\documentclass[sn-mathphys,Numbered]{sn-jnl}

\usepackage{graphicx}%
\usepackage{multirow}%
\usepackage{amsmath,amssymb,amsfonts}%
\usepackage{amsthm}%
\usepackage{mathrsfs}%
\usepackage{mathptmx}
\usepackage[title]{appendix}%
\usepackage{xcolor}%
\usepackage{textcomp}%
\usepackage{manyfoot}%
\usepackage{booktabs}%
\usepackage{algorithm}%
\usepackage{algorithmicx}%
\usepackage{algpseudocode}%
\usepackage{listings}%
\usepackage{gensymb}
\usepackage{siunitx} 
\usepackage{chemformula} 
\usepackage{pdfpages}
\usepackage{natbib}
\usepackage{lmodern}
\begin{document}

\title[Article Title]{A strongly interacting, two-dimensional, dipolar spin ensemble in (111)-oriented diamond}

\author[1]{\fnm{Lillian B.} \sur{Hughes}}\email{lbhughes@ucsb.edu}

\author[2]{\fnm{Simon A.} \sur{Meynell}}\email{simonmeynell@physics.ucsb.edu}

\author[3]{\fnm{Weijie} \sur{Wu}}\email{weijiewu@g.harvard.edu}

\author[2]{\fnm{Shreyas} \sur{Parthasarathy}}\email{sparthasarathy@ucsb.edu}

\author[2]{\fnm{Lingjie} \sur{Chen}}\email{lingjiechen@physics.ucsb.edu}

\author[2]{\fnm{Zhiran} \sur{Zhang}}\email{zhiranzhang@ucsb.edu}

\author[3]{\fnm{Zilin} \sur{Wang}}\email{zilinwang@g.harvard.edu}

\author[4]{\fnm{Emily J.} \sur{Davis}}\email{ejd8781@nyu.edu}

\author[5]{\fnm{Kunal} \sur{Mukherjee}}\email{kunalm@stanford.edu}

\author[3]{\fnm{Norman Y.} \sur{Yao}}\email{nyao@fas.harvard.edu}

\author*[2]{\fnm{Ania C.} \sur{Bleszynski Jayich}}\email{ania@physics.ucsb.edu}

\affil[1]{\orgdiv{Materials Department}, \orgname{University of California Santa Barbara}, \orgaddress{\city{Santa Barbara}, \postcode{93106}, \state{CA}, \country{U.S.A}}}

\affil*[2]{\orgdiv{Department of Physics}, \orgname{University of California Santa Barbara}, \orgaddress{\city{Santa Barbara}, \postcode{93106}, \state{CA}, \country{U.S.A}}}

\affil[3]{\orgdiv{Department of Physics}, \orgname{Harvard University}, \orgaddress{\city{Cambridge}, \postcode{01451}, \state{MA}, \country{U.S.A}}}

\affil[4]{\orgdiv{Department of Physics}, \orgname{New York University}, \orgaddress{\city{New York}, \postcode{10012}, \state{NY}, \country{U.S.A}}}

\affil[5]{\orgdiv{Department of Materials Science and Engineering}, \orgname{Stanford University}, \city{Palo Alto}, \postcode{94305}, \state{CA}, \country{U.S.A}}

\abstract{
Systems of spins with strong dipolar interactions and controlled dimensionality enable new explorations in quantum sensing and simulation.
In this work, we investigate the creation of strong dipolar interactions in a two-dimensional ensemble of nitrogen-vacancy (NV) centers generated via plasma-enhanced chemical vapor deposition (PECVD) on (111)-oriented diamond substrates.
We find that diamond growth on the (111) plane yields high incorporation of spins, both nitrogen and NV centers, where the density of the latter is tunable via the miscut of the diamond substrate. 
Our process allows us to form dense, preferentially aligned, 2D NV ensembles with volume-normalized AC sensitivity down to $\eta_{AC}$ = 810 pT µm$^{3/2}$ Hz$^{-1/2}$.
Furthermore, we show that (111) affords maximally positive dipolar interactions amongst a 2D NV ensemble, which is crucial for leveraging dipolar-driven entanglement schemes and exploring new interacting spin physics.
}




\maketitle
\section*{Main}
Defect spins in diamond such as the nitrogen-vacancy (NV) center have become an important platform for quantum technologies. 
Notable developments with single NVs include nanoscale sensing of condensed matter ~\cite{Casola2014,Tetienne2014} and biological systems~\cite{Barry2016}, as well as tests of entanglement ~\cite{Dolde2013,Hensen2015,Bradley2019, Hanson2019,Xie2021}.
Recently, there has been increased interest in exploring dense ensembles of NVs, both to enhance the sensitivity in quantum metrology applications and also as a platform for quantum simulation~\cite{Zhou2020}.
In principle, the strong dipolar interactions between solid-state spins should enable the use of entanglement-enhanced metrology protocols, leading to improvements in sensitivity scaling beyond the so-called standard quantum limit; however, realizing these enhancements has remained an outstanding challenge in solid-state systems~\cite{Choi2017b, Block2023}.
In addition, both the equilibrium phase diagram as well as the non-equilibrium dynamics of positionally-disordered dipoles have remained  important open questions~\cite{reich1987glassy,ghosh2002coherent,rodriguez2010study}.
 
To date, the majority of experiments with NV ensembles have studied three-dimensional spin systems.
However, the exploration of two-dimensional systems, where the inter-spin spacing is larger than the dopant layer thickness, yields several advantages.
For quantum sensing, two-dimensional systems promise an enhanced spatial resolution (set by the surface proximity of the layer) as well as a better interaction-induced decoherence scaling~\cite{Hughes2023,Davis2023}.
From the perspective of quantum simulation, lower dimensional systems  are crucial for investigating phenomena such as interaction-driven localization~\cite{Yao2014,Burin2015,Abanin2019} and for stabilizing topological phases such as quantum spin liquids~\cite{yao2018quantum,zou2017frustrated}. 
In 3D systems, the angular dependence of the dipolar interaction causes it to average to zero, leading to another broad obstruction for exploring dipolar-driven, many-body physics.
To this end, in recent years, there has been a tremendous amount of interest and progress in generating 2D spin systems (NV and nitrogen) in diamond via delta-doping during plasma-enhanced chemical vapor deposition (PECVD) growth on the (001) crystallographic plane~\cite{Hughes2023}.
These techniques have already facilitated the exploration of the decoherence dynamics of  two-dimensional NV spin ensembles~\cite{Davis2023}.

Despite this advancement, limitations related to (001) NV systems leave room for further improvement.
On the practical side, impurity addition during (001) PECVD diamond growth is known to increase surface roughening~\cite{Tokuda2007} and step bunching~\cite{deTheije2000,Yamada2015}, which promotes inhomogeneous dopant aggregation and compromises well-defined dimensionality and density.
Furthermore, many studies show that nitrogen incorporation efficiency is limited to $\sim$10-30 ppm ($\sim2\times$10$^{18}$ atoms/cm$^{3}$) in CVD-grown (001) films~\cite{Watanabe2009,Khomich2015,Eichhorn2019,Tallaire2020}.
NV incorporation during (001) growth is even lower than substitutional nitrogen, requiring enhancement techniques such as irradiation and annealing that may introduce vacancy-related damage and disorder.
Perhaps most fundamentally, an unfortunate consequence of the tetrahedral symmetry of diamond is that even for a 2D layer of NVs in (001)-oriented diamond, the dipolar interactions also exactly average to zero (see SI section 1).

A natural solution to these challenges is to investigate NV ensemble creation on another crystallographic plane. 
Growth on (111)-oriented diamond offers known advantages such as increased dopant incorporation~\cite{Samlenski1995} and preferential alignment of NVs along the $\langle 111 \rangle$ direction~\cite{Lesik2014,Michl2014,Fukui2014,Miyazaki2014,Ozawa2017, Osterkamp2019}, which improves measurement contrast as background fluorescence from other NV orientations is suppressed.
However, compared to (001), growth on the (111) diamond plane is largely under-explored, primarily due to minimal commercial access to substrates with reasonable size and expense, the difficulty of mechanical polishing, and conventional understanding that (111) epitaxy is plagued by extended defects such as twins, dislocations, and stacking faults~\cite{Sakaguchi1999,Kasu2003,Tokuda2008}.
Here, we use polished, type IIa diamond (111) substrates and employ a gentle PECVD growth condition (750 W, 0.05-0.1$\%$ $^{12}$CH$_4$) to achieve high-quality diamond epitaxy.

\begin{figure*} [hbt!]
\centering
\includegraphics [width = 130 mm] {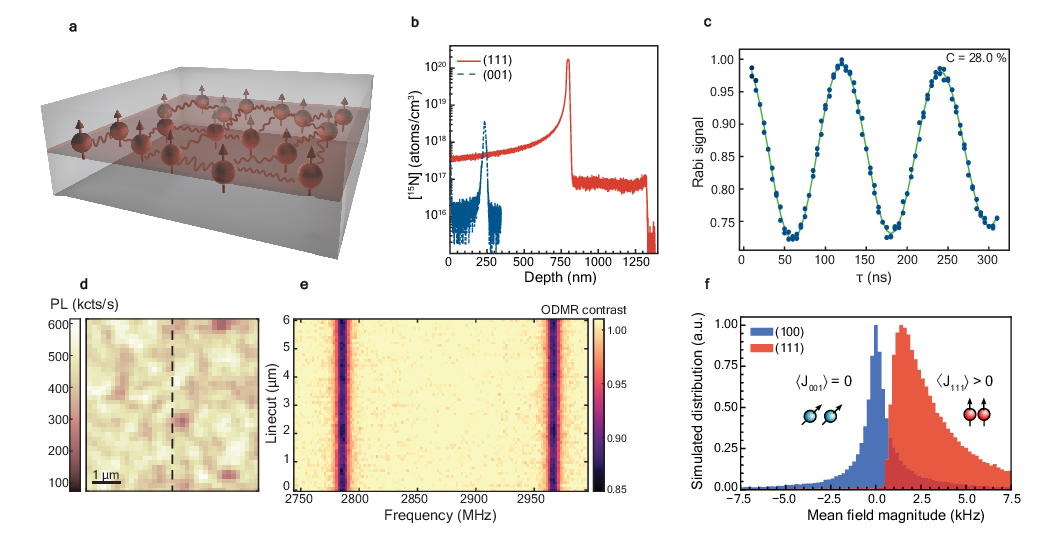}
\caption{\label{fig:fig1} \textbf{Creation of dense, 2D, (111) NV ensembles.} (a) Schematic illustrating a preferentially aligned (111) 2D NV ensemble with strong NV-NV interactions. 
(b) Secondary ion mass spectrometry (SIMS) depth profile showing higher nitrogen concentration in a (111) delta-doped diamond (red, solid - later referred to as sample C) as compared to an (001) sample (blue, dashed). Both substrates have similar miscut angles of ($\sim$1.6$\degree$) and were grown with similar doping conditions: 10 minute doping time and a gas ratio of [$^{15}$N$_2$]/[$^{12}$CH$_4$] = 12.5.
(c) Rabi measurement on a (111) 2D NV ensemble (later referred to as sample A) showing high contrast indicative of the preferential alignment. 
(d) PL scan of sample A. The dashed line shows a linecut over which ODMR is performed in the presence of a $\sim$32 Gauss field. 
(e) The ODMR data show preferential alignment of the grown-in NV centers along the $\langle$111$\rangle$ direction, with no other NV orientations visible at any point in the diamond epilayer.
(f) Simulated dipolar interaction-induced mean field distribution for polarized, 17 nm thick (001) (blue) and (111) (red) NV ensembles with a single-group density of 2 ppm$\cdot$nm, illustrating an asymmetric distribution for (111) due to the uniformly positive dipolar interaction energy.}
\end{figure*}

In this work, we demonstrate the creation of strongly interacting, 2D NV ensembles via nitrogen delta-doping during (111) diamond growth (see illustration in Fig.~\ref{fig:fig1}a).
We first investigate the mechanisms of growth and dopant incorporation on the (111) plane and find, as shown in Fig.~\ref{fig:fig1}b, that nitrogen incorporation is up to $\sim$60 times greater in (111) than (001).
We also explore NV creation during diamond growth, finding that the substrate miscut influences the density of NV centers formed.
We therefore use high miscut angle as a means of generating dense (up to 5.5 ppm$\cdot$nm), 2D NV ensembles with high contrast due to preferential alignment (shown by Fig.~\ref{fig:fig1}c-e), and volume-normalized sensitivity down to $\eta_{AC}$ = 810 pT µm$^{3/2}$ Hz$^{-1/2}$.
Lastly, we develop a novel protocol to characterize the non-zero dipolar interactions that are maximized in a 2D (111) NV ensemble. 
The uniformly positive interaction distribution, illustrated by the mean field simulation in Fig.~\ref{fig:fig1}f, results in an asymmetric spin resonance lineshape that we experimentally confirm in a (111) NV ensemble.
Altogether, this work demonstrates novel control over NV dipolar interactions that is enabled by (111) diamond growth. 

\subsection*{(111) Diamond growth and nitrogen delta-doping}
\begin{figure*}[hbt!]
\centering
\includegraphics[scale = 0.95] {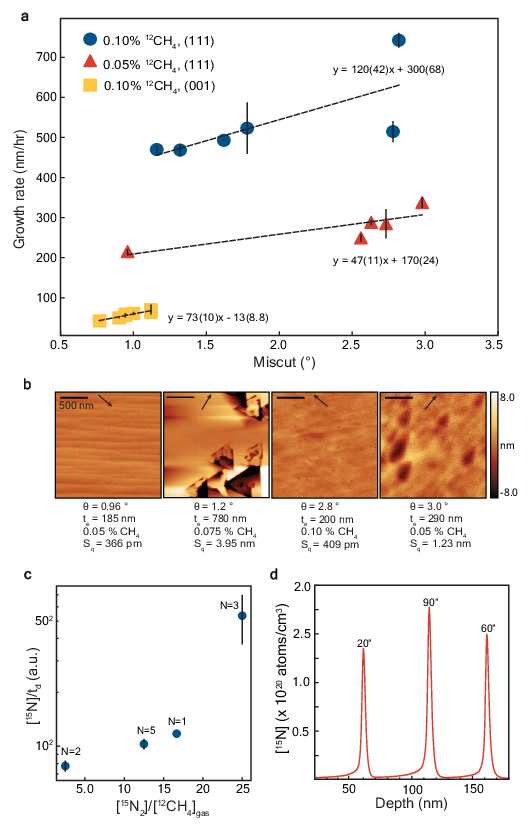}
\caption{\label{fig:fig2} \textbf{(111) PECVD growth, surface morphology, and delta-doped nitrogen incorporation.} (a) Diamond growth rate versus substrate miscut angle. Blue circles show (111) samples grown with 0.1\% CH$_{4}$, orange triangles show (111) with 0.05\% CH$_{4}$, and yellow shows (001) with 0.1\% CH$_{4}$. Error bars are estimated from the $^{13}$C SIMS step function used to calculate growth rate of the epitaxial layer. Trend lines are the result of a weighted linear regression and include the standard error. (b) AFM showing growth morphology with various polar miscut angles ($\theta$), epilayer thickness (t$_e$), methane content in the growth plasma, and resulting rms roughness (S$_q$). The azimuthal miscut direction is indicated by inset arrows. The highest and lowest miscut samples were grown simultaneously and are later referenced in this work as sample A and B, respectively. (c) Plot of SIMS-measured nitrogen concentration (normalized by doping time, t$_d$) versus gas flow ratios showing increasing incorporation with higher nitrogen content in the growth plasma. Data points represent an average of N diamond samples and error bars show the standard error of the mean. Our highest measured 2D density (sample C, 1.9(0.2)$\times10^{4}$ ppm$\cdot$nm) was grown with a [$^{15}$N$_{2}$]/[$^{12}$CH$_{4}$] ratio of 12 and doping time of 10 minutes. (d) SIMS profile of a sample grown with three different doping times, as indicated on the plot. The data show thin layer confinement (FWHM $\approx$ 4 nm for each peak) and slight increased density with longer doping time.}
\end{figure*}

Creating spin ensembles in diamond with control over density, dimensionality, and coherence relies crucially on understanding the mechanisms of growth and dopant incorporation.
In Fig.~\ref{fig:fig2}a, we explore the effects of substrate miscut and methane concentration on PECVD growth, finding that the (111) growth rate is several times faster than (001) (seen in the comparison of (001) rates (yellow squares) versus (111) rates (blue circles and orange triangles).
Faster growth on the (111) plane aligns with the models proposed in ref.~\cite{Larsson2015}, which suggest a higher rate of hydrogen abstraction on the (111) surface. 
The miscut-dependent (111) trends in Fig.~\ref{fig:fig2}a suggest two growth mechanisms at play: firstly, the non-zero y-intercept indicates the presence of a miscut-independent growth mechanism and secondly, the non-zero slope suggests an additional contribution from a step-edge-dependent growth mode.
Both the slope and intercept of the (111) data approximately double with the carbon content in the gas: 0.05\% CH$_{4}$ (red triangles) versus 0.10\% CH$_{4}$ (blue circles) in Fig.~\ref{fig:fig2}a.
This methane concentration dependence implies that the (111) growth is adatom-limited and that carbon addition is not restricted only to step-edge sites, further supporting that multiple growth mechanisms are at play.
In contrast, the nearly zero intercept of the (001) trend indicates that our (001) growth is dominated by step-flow, supporting previous observations\cite{Meynell2020}.
Atomic force microscopy (AFM) analysis shows a variety of surface morphologies mediated by substrate miscut (both polar angle, $\theta$, and azimuthal), CH$_{4}$ content in the growth plasma, and epilayer thickness (t$_e$) (Fig.~\ref{fig:fig2}b).
Overall, the morphologies are suitable for our applications as the surface roughness is still $\lesssim$ 1 nm in most samples. 

We next investigate control over nitrogen incorporation during (111) growth and demonstrate confinement of high-density nitrogen into thin delta-doped layers. 
As shown in Fig.~\ref{fig:fig1}b, nitrogen incorporation is enhanced by a factor of $\sim$60 on the (111) plane as compared to our highest density (001) sample grown with the same doping conditions.
The (111) sample has an areal density of 1.9(0.2)$\times10^{4}$ ppm$\cdot$nm (3.43(0.04)$\times10^{14}$ atoms/cm$^{2}$), as determined from integrating the secondary ion mass spectrometry (SIMS) data as in ref.~\cite{Hughes2023}.
Increased dopant incorporation on the (111) plane is consistent with observations from three-dimensional doping studies~\cite{Kato2007,Samlenski1995,Mortet2022}; however, we note that these works observe a smaller enhancement of only 3-4x.
We find that our main control knobs over nitrogen density are the ratio of the source gases ([$^{15}$N$_{2}$]/[$^{12}$CH$_{4}$]) and the doping time (t$_d$).
Fig.~\ref{fig:fig2}c shows nitrogen density, as determined with SIMS and normalized by t$_d$, versus ([$^{15}$N$_{2}$]/[$^{12}$CH$_{4}$]), highlighting greater incorporation with higher nitrogen concentration in the plasma.
We do not increase [$^{15}$N$_{2}$]/[$^{12}$CH$_{4}$] beyond 25 due to the limitations of our mass flow controller (we note that 25 is already larger than typical ratios~\cite{Kato2007, Nakano2022}), but future studies will explore a higher range.  
Notably, Fig.~\ref{fig:fig2}c indicates that we have not reached the surface solubility limit for the (111) diamond surface, in contrast to observations in (001) diamond~\cite{Watanabe2009}, leaving open the possibility of achieving even higher nitrogen densities in (111). 

The thickness of the nitrogen layers can be reduced to a few nanometers for sufficiently short doping time; 
Fig.~\ref{fig:fig2}d shows three thin nitrogen layers in a single sample with doping times of 20, 60, and 90 seconds. 
Each layer is measured to be 4 nm thick, which gives an upper bound as the measured thickness is limited by the SIMS resolution in this case.
Even for this overestimate, the two-dimensional nature holds as the spacing between spins is larger than the layer thickness. 
Lengthening the doping time allows us to increase the nitrogen areal density while staying within the 2D limit, which may be afforded in part by a decreased growth rate during doping due to the high nitrogen content in the plasma (see SI section 2).
Even longer doping times lead to 3D spin systems. 
The relevant dimensionality of the dipolar interacting spin system can be probed via the decoherence dynamics of the spins themselves as in ref.~\cite{Hughes2023,Davis2023}, wherein the stretch exponent of the coherence decay indicates the dimensionality of the spin bath. 
We apply this technique to verify the 2D nature of the dipolar spin bath in our (111) delta-doped samples, as shown later in this work.
Lastly, we find that in contrast to (001) growth~\cite{Tallaire2015}, (111) nitrogen incorporation shows a slight increase with higher doping temperature (similar to ref.~\cite{Kato2007}) and no significant dependence on the substrate miscut, also unlike (001) growth~\cite{Meynell2020}.
SI section 2 provides further discussion on the effects of temperature and miscut on nitrogen incorporation in (111). 

\subsection*{Creation of dense, preferentially aligned 2D NV ensembles}\label{subsec3}
We next explore NV center incorporation during (111) diamond growth.
Although preferential alignment of NVs during (111) growth has been previously studied, here we focus on a new regime targeting high densities (where NV-NV dipolar interactions dominate) and two-dimensionality, which has not previously been demonstrated in (111) diamond.
Prior studies investigate ppb-level NV incorporation that originates from low-concentration doping or impurity nitrogen in the hydrogen and methane source gases~\cite{Ozawa2017,Osterkamp2019}.
Ref.~\cite{Ishiwata2017} observes NV densities up to 0.18 ppm (3.1$\times$10$^{16}$ atoms/cm$^{3}$) as-grown in (111) diamond, but they are determined via PL intensity, which is prone to overestimation due to non-NV$^{-}$ fluorescent defects.
Traditional methods to enhance the NV density include irradiation and annealing treatments, which come at the cost of reduced preferential alignment.
A method to generate high-density, strongly interacting NV ensembles while preserving (111) alignment has not yet been demonstrated until this work.

\begin{figure}[hbt!]
\centering
\includegraphics [scale = 0.95] {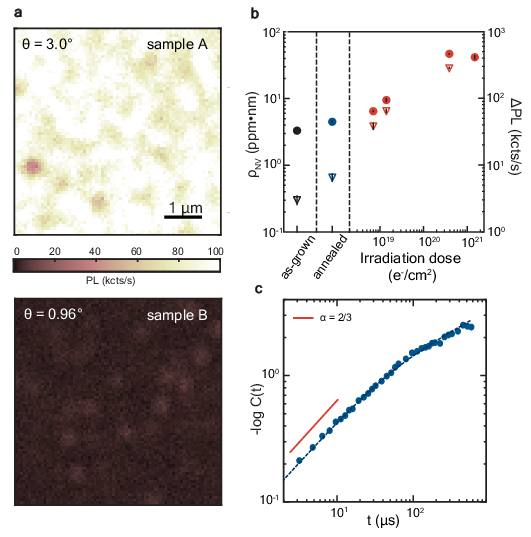}
\caption{\label{fig:fig3} \textbf{Miscut as a tuning knob for NV incorporation during growth.} (a) As-grown scanning confocal PL images taken under 80 µW excitation for samples A and B grown simultaneously with different miscuts of 3.0$\degree$ and 0.96$\degree$, respectively. The color scale is the same for the two images, showing increased NV fluorescence in sample A (higher miscut). (b) Plot of NV density in sample A (circles) and B (triangles) during different stages of processing. The dashed line highlights different stages in the processing timeline, from as-grown to post-electron irradiation (red data). The as-grown density of sample A, $\rho_{NV}$ = 4.5 $\pm$ 0.5 ppm$\cdot$n,was measured at three different locations on the sample with the decoherence based method (the saturated coherence of the NV ensemble under an XY-8 sequence) and calibrated to an optical measurement of $\Delta$PL, a proxy for NV density that represents the intensity difference between the $m_s=0$ and $\pm1$ states in a Rabi experiment as described in ref.~\cite{Hughes2023}. All other data points were measured via $\Delta$PL, plotted on the right-side y-axis of Fig.~\ref{fig:fig3}b, with the left-side y-axis linearly scaled. (c) Log-log plot of the NV XY-8 decay in sample A, highlighting the stretch exponent power $\alpha$ = 2/3 in the early-time regime.}
\end{figure}

We find that the substrate miscut plays a surprisingly large role in the concentration of aligned NV centers that incorporate during diamond growth, in stark contrast to the observed lack of dependence of nitrogen incorporation on miscut.
Fig.~\ref{fig:fig3}a shows scanning confocal PL images for samples A and B, grown simultaneously with a doping time of one minute and differing only in substrate miscut of 3.0$\degree$ and 0.96$\degree$, respectively.
The images are taken immediately post-growth and show clear increased PL from sample A.
Furthermore, all of the NVs display preferential alignment, as shown in the ODMR spectra of sample A in Fig.~\ref{fig:fig1}c.
We quantitatively estimate the grown-in, aligned NV density in sample A using a decoherence-based method as described in ref.~\cite{Hughes2023} and extract an average density of $\rho_{NV}$ = 4.5 $\pm$ 0.5 ppm$\cdot$nm.
Without irradiation, this density is already similar to the highest NV densities we have previously achieved via irradiation and annealing of (001) samples~\cite{Hughes2023,Davis2023}. 
Fig.~\ref{fig:fig3}b shows how NV concentration in sample A and B evolves with irradiation and annealing, achieving densities up to 47 ppm$\cdot$nm and 28 ppm$\cdot$nm, respectively.
The NV density in sample B approaches a similar level to sample A after irradiation, consistent with the comparable nitrogen content seen via SIMS in the two samples (SI section 3).
We note that the trend of increasing grown-in NV density with miscut angle holds for all additional samples measured (SI section 4).

Lastly, we discuss the strongly interacting, two-dimensional nature of the high-density NV layers. 
By measuring the stretch exponent $n$ of the XY-8 coherence decay, $e^{-(t/T_2)^n}$, we find $n = 2/3$ as expected for a 2D dipolar-interacting ensemble at early times (where t $<\tau_c$, the correlation time of the ensemble)~\cite{Davis2023} (Fig.~\ref{fig:fig3}c).
The dimensionality of the nitrogen is further corroborated by an $n=2/3$ stretch exponent in the double electron-electron resonance (DEER) decay, which probes neutrally-charged substitutional nitrogen, [N$_s^0$] or P1 centers (SI section 3). 
With DEER we also extract the density of P1 centers as $\rho_{P1}$ = 130 $\pm$ 19 ppm$\cdot$nm. 
Dividing $\rho_{NV}$/$\rho_{P1}$ gives the as-grown NV yield in sample A as $\sim4$\%.
Thus, by using miscut as a tuning knob during growth, we can achieve a strongly interacting, aligned 2D NV ensemble where NV dipolar interactions dominate due to the high ratio of NV to nitrogen.

\subsection*{Magnetic sensing with (111) NV ensembles}
With the ability to form aligned, dense ensembles at the dipolar limit, we now investigate their applicability to ensemble sensing applications and quantum simulation. 
We first discuss the AC magnetic sensitivity of our 2D (111) ensembles.
We arrive at a measured, volume-normalized AC sensitivity of $\eta_{AC}$ = 810 pT µm$^{3/2}$ Hz$^{-1/2}$ for an electron-irradiated spot on sample B (7 $\times$10$^{18}$ e$^-$/cm$^{2}$). 
With reasonable improvements to photon collection efficiency, namely count rates of 150 kCts/sec per NV as can be achieved with e.g., nanopillars, and all other parameters identical to those measured on our setup on this sample, we project $\eta_{AC}$ = 153 pT µm$^{3/2}$ Hz$^{-1/2}$. 
The volume-normalized sensitivities we obtain are improvements over prior reports of NV ensembles in 3D (001) samples~\cite{Zhou2020,Eichhorn2019,Wolf2015} and 3D (111) samples~\cite{Osterkamp2020}. 
Further improvements to sensitivity can be gained by applying dipolar-decoupling sequences such as DROID-60 ~\cite{Zhou2020}. 
In SI section 5, we discuss the sensitivity calculations and also demonstrate magnetic-field imaging of a Fe nanoparticle using our (111) NV ensembles. 
Finally, we note that future dipolar-driven entanglement-enhanced sensing schemes~\cite{Block2023,Perlin2020,Bornet2023} may depend crucially on dense, 2D (111) ensembles in which dipolar interactions do not average to zero over an ensemble, as we confirm in the next section.

\subsection*{Characterization of the (111) NV-NV dipolar interaction}
\begin{figure}[hbt!]
\centering
\includegraphics[scale = 0.95] {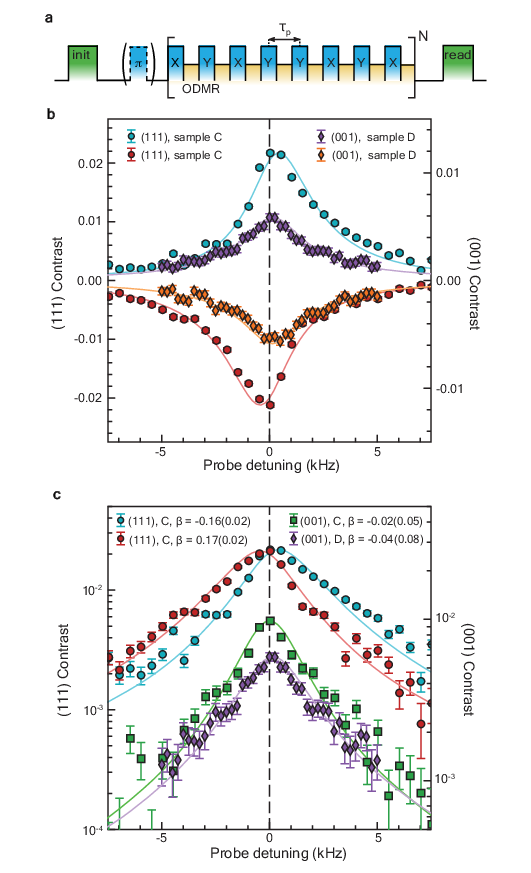}
\caption{\label{fig:fig4} \textbf{Disorder-suppressed, dipolar-limited ODMR spectra of NV ensembles.} (a) XY8-ODMR pulse sequence schematic: NVs are initialized into either a bright ($m_s = 0$) state via a 532 nm laser pulse, or dark ($m_s = \pm 1$) state via an additional strong $\pi$ pulse. (b) XY8-ODMR spectra, normalized as a fraction of total PL, for NV ensembles in (111)-oriented sample C and (001)-oriented sample D. Purple (orange) diamonds are measured spectra for Sample D for initialization into the bright (dark) states and cyan (red) circles are measured spectra for Sample C for initialization into the bright (dark) states. 
Solid lines are simulations of the mean field distribution as shown in Fig.~\ref{fig:fig1}e, incorporating various broadening mechanisms (see SI section 7). (c) Spectra on a log-linear scale to highlight the spin orientation-dependent asymmetry in the (111) NV ensemble (red and cyan circles, sample C). The purple (001) sample D XY8-ODMR spectrum from (b) is reproduced for comparison. To demonstrate that the asymmetry is a fundamental property of (111)-oriented ensembles, we also display the dipolar-limited spectrum of other NV orientations in sample C (green squares). $\beta$ is defined as the difference in normalized mean area under the curve between the left and right side of each spectrum and emphasizes the asymmetry of the (111) data.
}
\end{figure}

Lastly, we present direct characterization of the uniformly positive dipolar interaction unique to a (111) 2D NV ensemble, which is observed as an asymmetric lineshape in an ODMR measurement.
Fig.~\ref{fig:fig1}e shows simulated dipolar-induced mean field distributions for (001) and (111) 2D NV ensembles using the experimentally determined density and thickness (SI section 6), illustrating the non-zero average interactions present in (111).
To experimentally access the dipolar-limited lineshape of this distribution, we apply a new technique, here termed XY8-ODMR, which is designed to suppress disorder during ODMR. 
This measurement first uses an XY8-style ``pump" sequence (shown in Fig.~\ref{fig:fig4}a, details in Methods and SI section 7) to resonantly drive and decouple the NV ensemble from sources of disorder in the surrounding lattice.
Under this drive, the remaining sources of broadening for the NV ensemble lineshape are dominated by dipolar interactions, which are not affected by the pump sequence.
Simultaneous to the XY-8 drive, we deliver a weak ``probe" RF $\pi$ pulse, analogous to the $\pi$ pulse used in a conventional pulsed ODMR sequence ~\cite{Dreau2011}.
By varying the frequency of the probe $\pi$ pulse and monitoring the resulting spin population, we obtain a disorder-suppressed equivalent of a pulsed ODMR spectrum (an ``XY8-ODMR" spectrum) that is not linewidth-limited by disorder (e.g., due to inhomogeneous strain or coupling to other defects) but rather primarily by dipolar interactions and power broadening.
In this regime, the measured XY8-ODMR spectrum lineshape is a direct measure of the distribution of dipolar couplings in the ensemble. 

Fig.~\ref{fig:fig4}b-c shows an asymmetric lineshape for a (111) NV ensemble (sample C), compared to a symmetric lineshape for (001) (sample D). 
We also confirm that the other NV orientations (not normal to the 2D plane, as in an (001)-oriented diamond) in sample C exhibit a symmetric lineshape, as expected. 
The sign of the asymmetry observed in the (111) ensemble flips when the spins are initialized into the opposite spin state, demonstrating that the source of the mean field is the spins themselves. 
Altogether, the asymmetry observed here highlights a signature of the uniformly positive dipolar interactions in (111) that are crucial for further explorations in entanglement-enhanced metrology and quantum simulation.

\section*{Outlook}
Looking forward, we expect the results presented in this work to inspire widespread implementation of delta-doped (111) diamond into sensing and simulation experiments, in particular leveraging the unique form of the dipolar interactions in 2D (111) ensembles for entanglement-enabled enhancements. 
We anticipate that the high sensitivities demonstrated already here will be exploited for biological and condensed matter sensing, with the potential of increased sensitivity through more advanced dynamical decoupling sequences such as DROID-60~\cite{Zhou2020}.
Further materials engineering directions include pushing to higher nitrogen and NV densities by, for instance, increasing the source gas nitrogen content ([$^{15}$N$_{2}$]/[$^{12}$CH$_{4}$]) beyond what we explore here or optimizing post-growth processing with higher dose irradiation or higher temperature annealing (e.g., 1200 $\degree$C).
Another frontier to explore is improving sensitivity and spatial resolution by bringing (111) NV ensembles close to the diamond surface.
This direction requires improved understanding of the (111) diamond surface, both in morphology (roughness, sources of magnetic noise) and termination (charge stability), in order to engineer coherent and stable shallow NVs.~\cite{Chou2017, Gali2017}
It will be important to understand the role of miscut, specifically the azimuthal angle, and growth conditions (i.e., achieving wholly lateral growth) on surface morphology. 
The potential for growth on atomically smooth (111) surfaces~\cite{Tokuda2008,Parks2018} is intriguing for even higher quality growth. 
All in all, this work lays an important foundation for diamond quantum technology and opens several avenues for further exploration.

\section*{Methods}
\subsection*{PECVD diamond growth}
Diamond substrates are obtained by slicing (001)-oriented electronic grade CVD substrates (Element Six Ltd.) along the (111) plane and super polishing by Applied Diamond, Inc. to achieve a surface roughness of $<$ 500 pm. 
Prior to growth, the miscut angle is measured using x-ray diffractometry (XRD) rocking curves about the (111) omega peak and surface roughness is verified with AFM.
PECVD growth is performed in a SEKI SDS6300 reactor using a 750 W plasma containing 400 sccm H$_2$ and 0.05-0.1$\%$ $^{12}$CH$_{4}$ held at 25 torr and $\sim$770 $^{\circ}$C according to a pyrometer.
These growth conditions result in a 99.998$\%$ $^{12}$C isotopically purified epilayer of variable thickness dependent on the growth time and substrate miscut.
During the nitrogen delta-doping period of growth, $^{15}$N$_2$ gas (1.25$\%$ of the total gas content) is introduced into the chamber for a variable doping time of 0.5-10 minutes. 
Optical emission spectroscopy (OES) is used to monitor the plasma species throughout growth, specifically the H-alpha emission line and CN radicals.
After growth, SIMS is used to estimate the isotopic purity, epilayer thickness, and properties of the delta-doped layer.
SIMS analysis is performed in-house with conditions as outlined previously in ref.~\cite{Hughes2023}.

\subsection*{Post-growth processing}
Electron irradiation is performed with a transmission electron microscope (TEM, ThermoFisher Talos F200X G2) to create locally enhanced NV concentrations inside the irradiated region~\cite{McLellan2016} .
Afterwards, the samples undergo subsequent annealing at $400 \degree$ C for 2 hours and $850 \degree$ C for 4 hours in a vacuum furnace (4e-8 torr) to promote vacancy diffusion and NV formation. 
After irradiation and annealing, the samples are cleaned in a boiling tri-acid solution (1:1:1 H$_2$SO$_4$:H$_2$O$_2$:HClO$_4$) and annealed at $450 \degree$ C for 4 hours in air to remove surface contaminants and promote oxygen termination which stabilizes the negative NV$^{-}$ charge state. 

\subsection*{NV center characterization}
All NV center measurements are performed on a home-built confocal microscope using a 532-nm diode laser and an external magnetic field of $\sim$300-370 G aligned along the diamond $\langle 111 \rangle$ axis unless otherwise specified.
A cutoff filter of 640 nm is placed in the collection path before the avalanche photodiode to eliminate background fluorescence from NV$^{0}$.
Radiofrequency (RF) signals are delivered through a gold antenna fabricated on a glass slide and placed underneath the diamond. 

\subsection*{XY8-ODMR protocol}
As depicted in Fig.~\ref{fig:fig4}a, the protocol for probing dipolar interactions starts with pumping the NV centers into the m$_s$ = 0 state with a pulse of 532 nm light (power $P_{\mathrm{laser}}$, duration $t_i$). 
To initialize the NV centers into the m$_s = \pm 1$ states, we use a resonant microwave $\pi$-pulse at a Rabi frequency of $\Omega_\mathrm{pump}$, drive frequency $f_\mathrm{NV}$. 
We use the same microwave source to perform dynamical decoupling and remove as much disorder as possible. 
For an XY8-N decoupling sequence, the $\pi$ pulse spacing $\tau_p$ was decreased (to shift the NV coherence filter function to higher frequencies) until the measured coherence under the dynamical decoupling drive saturated, indicating that disorder was suppressed as much as possible. 
Having set $\tau_p$ optimally, the probe low-power ODMR drive was swept across a center frequency $f_\mathrm{NV} + \frac{1}{4 \tau_p}$ with power $\Omega_\mathrm{probe}$. 
This offset center frequency is necessary to properly account for the observed modulated resonant frequency of NVs in the toggling frame of the XY8 drive (see SI section 7). 
The number of repetitions $N$ were set such that the length of the dynamical decoupling sequence $8 N\tau_p$ was equal to the length of the ODMR $\pi$-pulse $\frac{\pi}{\Omega_\mathrm{probe}}$. 
Finally, the spin population was read out with a 532nm laser pulse (duration $t_r$). 
The final sequence parameters used for each sample are:
\begin{itemize}
    \item Sample C: $t_i = 40\mu\mathrm{s}, \quad P_{\mathrm{laser}} = 100\mu\mathrm{W}, \quad \Omega_\mathrm{pump} = 2\pi \times 12.5$MHz, $\quad \tau_p = 540$ns, $\quad \Omega_\mathrm{probe} = 2\pi \times 0.97$kHz, $\quad t_r = 1000$ns
    \item Sample D: $t_i = 40\mu\mathrm{s}, \quad P_{\mathrm{laser}} = 180\mu\mathrm{W}, \quad \Omega_\mathrm{pump} = 2\pi \times 16$ MHz, $\quad \tau_p = 1.032\mu$s, $\quad \Omega_\mathrm{probe} = 2\pi \times 0.75$kHz, $\quad t_r = 400$ns
\end{itemize}

\bmhead{Supplementary information}
See the SI for additional details on the (111) nonzero dipolar interaction, controlling nitrogen incorporation, influence of miscut on NV creation during growth, sensitivity calculations and demonstration of ensemble magnetic imaging, additional sample information, and XY8-ODMR characterization of the dipolar-limited NV lineshape.

\bmhead{Acknowledgments}
We gratefully acknowledge support of the U.S. Department of Energy BES grant No. DE-SC0019241 (materials growth and characterization), the Center for Novel Pathways to Quantum Coherence in Materials, an Energy Frontier Research Center funded by the U.S. Department of Energy, Office of Science, Basic Energy Sciences (coherence studies), and the Army Research Office through the MURI program grant number W911NF-20-1-0136 (theoretical studies).
We acknowledge the use of shared facilities of the UCSB Quantum Foundry through Q-AMASE-i program (NSF DMR-1906325), the UCSB MRSEC (NSF DMR-2308708), and the Quantum Structures Facility within the UCSB California NanoSystems Institute.
L. B. H. acknowledges support from the NSF Graduate Research Fellowship Program (DGE 2139319) and the UCSB Quantum Foundry.
S. A. M. acknowledges support from Canada NSERC (AID 516704-2018) and the UCSB Quantum Foundry. 
S.P. acknowledges support from the Department of Defense (DoD) through the National Defense Science and Engineering Graduate (NDSEG) Fellowship Program.

\bibliography{111_Main}


\begin{thebibliography}{54}
\ifx \bisbn   \undefined \def \bisbn  #1{ISBN #1}\fi
\ifx \binits  \undefined \def \binits#1{#1}\fi
\ifx \bauthor  \undefined \def \bauthor#1{#1}\fi
\ifx \batitle  \undefined \def \batitle#1{#1}\fi
\ifx \bjtitle  \undefined \def \bjtitle#1{#1}\fi
\ifx \bvolume  \undefined \def \bvolume#1{\textbf{#1}}\fi
\ifx \byear  \undefined \def \byear#1{#1}\fi
\ifx \bissue  \undefined \def \bissue#1{#1}\fi
\ifx \bfpage  \undefined \def \bfpage#1{#1}\fi
\ifx \blpage  \undefined \def \blpage #1{#1}\fi
\ifx \burl  \undefined \def \burl#1{\textsf{#1}}\fi
\ifx \doiurl  \undefined \def \doiurl#1{\url{https://doi.org/#1}}\fi
\ifx \betal  \undefined \def \betal{\textit{et al.}}\fi
\ifx \binstitute  \undefined \def \binstitute#1{#1}\fi
\ifx \binstitutionaled  \undefined \def \binstitutionaled#1{#1}\fi
\ifx \bctitle  \undefined \def \bctitle#1{#1}\fi
\ifx \beditor  \undefined \def \beditor#1{#1}\fi
\ifx \bpublisher  \undefined \def \bpublisher#1{#1}\fi
\ifx \bbtitle  \undefined \def \bbtitle#1{#1}\fi
\ifx \bedition  \undefined \def \bedition#1{#1}\fi
\ifx \bseriesno  \undefined \def \bseriesno#1{#1}\fi
\ifx \blocation  \undefined \def \blocation#1{#1}\fi
\ifx \bsertitle  \undefined \def \bsertitle#1{#1}\fi
\ifx \bsnm \undefined \def \bsnm#1{#1}\fi
\ifx \bsuffix \undefined \def \bsuffix#1{#1}\fi
\ifx \bparticle \undefined \def \bparticle#1{#1}\fi
\ifx \barticle \undefined \def \barticle#1{#1}\fi
\bibcommenthead
\ifx \bconfdate \undefined \def \bconfdate #1{#1}\fi
\ifx \botherref \undefined \def \botherref #1{#1}\fi
\ifx \url \undefined \def \url#1{\textsf{#1}}\fi
\ifx \bchapter \undefined \def \bchapter#1{#1}\fi
\ifx \bbook \undefined \def \bbook#1{#1}\fi
\ifx \bcomment \undefined \def \bcomment#1{#1}\fi
\ifx \oauthor \undefined \def \oauthor#1{#1}\fi
\ifx \citeauthoryear \undefined \def \citeauthoryear#1{#1}\fi
\ifx \endbibitem  \undefined \def \endbibitem {}\fi
\ifx \bconflocation  \undefined \def \bconflocation#1{#1}\fi
\ifx \arxivurl  \undefined \def \arxivurl#1{\textsf{#1}}\fi
\csname PreBibitemsHook\endcsname

\bibitem[\protect\citeauthoryear{Casola et~al.}{2018}]{Casola2014}
\begin{botherref}
\oauthor{\bsnm{Casola}, \binits{F.}},
\oauthor{\bsnm{Sar}, \binits{T.}},
\oauthor{\bsnm{Yacoby}, \binits{A.}}:
Probing condensed matter physics with magnetometry based on nitrogen-vacancy
  centres in diamond.
Nat. Rev. Mater.
\textbf{3}(17088)
(2018)
\doiurl{10.1038/natrevmats.2017.88}
\end{botherref}
\endbibitem

\bibitem[\protect\citeauthoryear{Tetienne et~al.}{2014}]{Tetienne2014}
\begin{barticle}
\bauthor{\bsnm{Tetienne}, \binits{J.-P.}},
\bauthor{\bsnm{Hingant}, \binits{T.}},
\bauthor{\bsnm{Kim}, \binits{J.-V.}},
\bauthor{\bsnm{Herrera~Diez}, \binits{L.}},
\bauthor{\bsnm{Adam}, \binits{J.-P.}},
\bauthor{\bsnm{Garcia}, \binits{K.}},
\bauthor{\bsnm{Roch}, \binits{J.-F.}},
\bauthor{\bsnm{Rohart}, \binits{S.}},
\bauthor{\bsnm{Thiaville}, \binits{A.}},
\bauthor{\bsnm{Ravelosona}, \binits{D.}},
\bauthor{\bsnm{Jacques}, \binits{V.}}:
\batitle{Nanoscale imaging and control of domain-wall hopping with a
  nitrogen-vacancy center microscope}.
\bjtitle{Science}
\bvolume{344}(\bissue{6190}),
\bfpage{1366}--\blpage{1369}
(\byear{2014})
\doiurl{10.1126/science.1250113}
\end{barticle}
\endbibitem

\bibitem[\protect\citeauthoryear{Barry et~al.}{2016}]{Barry2016}
\begin{barticle}
\bauthor{\bsnm{Barry}, \binits{J.F.}},
\bauthor{\bsnm{Turner}, \binits{M.J.}},
\bauthor{\bsnm{Schloss}, \binits{J.M.}},
\bauthor{\bsnm{Glenn}, \binits{D.R.}},
\bauthor{\bsnm{Song}, \binits{Y.}},
\bauthor{\bsnm{Lukin}, \binits{M.D.}},
\bauthor{\bsnm{Park}, \binits{H.}},
\bauthor{\bsnm{Walsworth}, \binits{R.L.}}:
\batitle{Optical magnetic detection of single-neuron action potentials using
  quantum defects in diamond}.
\bjtitle{PNAS}
\bvolume{113}(\bissue{49}),
\bfpage{14133}--\blpage{14138}
(\byear{2016})
\doiurl{10.1073/pnas.1601513113}
\end{barticle}
\endbibitem

\bibitem[\protect\citeauthoryear{Dolde et~al.}{2013}]{Dolde2013}
\begin{barticle}
\bauthor{\bsnm{Dolde}, \binits{F.}},
\bauthor{\bsnm{Jakobi}, \binits{I.}},
\bauthor{\bsnm{Naydenov}, \binits{B.}},
\bauthor{\bsnm{Zhao}, \binits{N.}},
\bauthor{\bsnm{Pezzagna}, \binits{S.}},
\bauthor{\bsnm{Trautmann}, \binits{C.}},
\bauthor{\bsnm{Meijer}, \binits{J.}},
\bauthor{\bsnm{Neumann}, \binits{P.}},
\bauthor{\bsnm{Jelezko}, \binits{F.}},
\bauthor{\bsnm{Wrachtrup}, \binits{J.}}:
\batitle{Room-temperature entanglement between single defect spins in diamond}.
\bjtitle{Nat. Phys.}
\bvolume{9},
\bfpage{139}--\blpage{143}
(\byear{2013})
\doiurl{10.1038/nphys2545}
\end{barticle}
\endbibitem

\bibitem[\protect\citeauthoryear{Hensen et~al.}{2015}]{Hensen2015}
\begin{barticle}
\bauthor{\bsnm{Hensen}, \binits{B.}},
\bauthor{\bsnm{Bernien}, \binits{H.}},
\bauthor{\bsnm{Dréau}, \binits{A.E.}},
\bauthor{\bsnm{Reiserer}, \binits{A.}},
\bauthor{\bsnm{Kalb}, \binits{N.}},
\bauthor{\bsnm{Blok}, \binits{M.S.}},
\bauthor{\bsnm{Ruitenberg}, \binits{J.}},
\bauthor{\bsnm{Vermeulen}, \binits{R.F.L.}},
\bauthor{\bsnm{Schouten}, \binits{R.N.}},
\bauthor{\bsnm{Abellán}, \binits{C.}},
\bauthor{\bsnm{Amaya}, \binits{W.}},
\bauthor{\bsnm{Pruneri}, \binits{V.}},
\bauthor{\bsnm{Mitchell}, \binits{M.W.}},
\bauthor{\bsnm{Markham}, \binits{M.}},
\bauthor{\bsnm{Twitchen}, \binits{D.J.}},
\bauthor{\bsnm{Elkouss}, \binits{D.}},
\bauthor{\bsnm{Wehner}, \binits{S.}},
\bauthor{\bsnm{Taminiau}, \binits{T.H.}},
\bauthor{\bsnm{Hanson}, \binits{R.}}:
\batitle{Loophole-free bell inequality violation using electron spins separated
  by 1.3 kilometres}.
\bjtitle{Nature}
\bvolume{526},
\bfpage{682}--\blpage{686}
(\byear{2015})
\doiurl{10.1038/nature15759}
\end{barticle}
\endbibitem

\bibitem[\protect\citeauthoryear{Bradley et~al.}{2019}]{Bradley2019}
\begin{barticle}
\bauthor{\bsnm{Bradley}, \binits{C.E.}},
\bauthor{\bsnm{Randall}, \binits{J.}},
\bauthor{\bsnm{Abobeih}, \binits{M.H.}},
\bauthor{\bsnm{Berrevoets}, \binits{R.C.}},
\bauthor{\bsnm{Degen}, \binits{M.J.}},
\bauthor{\bsnm{Bakker}, \binits{M.A.}},
\bauthor{\bsnm{Markham}, \binits{M.}},
\bauthor{\bsnm{Twitchen}, \binits{D.J.}},
\bauthor{\bsnm{Taminiau}, \binits{T.H.}}:
\batitle{A ten-qubit solid-state spin register with quantum memory up to one
  minute}.
\bjtitle{Phys. Rev. X}
\bvolume{9},
\bfpage{031045}
(\byear{2019})
\doiurl{10.1103/PhysRevX.9.031045}
\end{barticle}
\endbibitem

\bibitem[\protect\citeauthoryear{van Dam et~al.}{2019}]{Hanson2019}
\begin{barticle}
\bauthor{\bsnm{Dam}, \binits{S.B.}},
\bauthor{\bsnm{Cramer}, \binits{J.}},
\bauthor{\bsnm{Taminiau}, \binits{T.H.}},
\bauthor{\bsnm{Hanson}, \binits{R.}}:
\batitle{Multipartite entanglement generation and contextuality tests using
  nondestructive three-qubit parity measurements}.
\bjtitle{Phys. Rev. Lett.}
\bvolume{123},
\bfpage{050401}
(\byear{2019})
\doiurl{10.1103/PhysRevLett.123.050401}
\end{barticle}
\endbibitem

\bibitem[\protect\citeauthoryear{Xie et~al.}{2021}]{Xie2021}
\begin{barticle}
\bauthor{\bsnm{Xie}, \binits{T.}},
\bauthor{\bsnm{Zhao}, \binits{Z.}},
\bauthor{\bsnm{Kong}, \binits{X.}},
\bauthor{\bsnm{Ma}, \binits{W.}},
\bauthor{\bsnm{Wang}, \binits{M.}},
\bauthor{\bsnm{Ye}, \binits{X.}},
\bauthor{\bsnm{Yu}, \binits{P.}},
\bauthor{\bsnm{Yang}, \binits{Z.}},
\bauthor{\bsnm{Xu}, \binits{S.}},
\bauthor{\bsnm{Wang}, \binits{P.}},
\bauthor{\bsnm{Wang}, \binits{Y.}},
\bauthor{\bsnm{Shi}, \binits{F.}},
\bauthor{\bsnm{Du}, \binits{J.}}:
\batitle{Beating the standard quantum limit under ambient conditions with
  solid-state spins}.
\bjtitle{Sci. Adv.}
\bvolume{7}(\bissue{32}),
\bfpage{9204}
(\byear{2021})
\doiurl{10.1126/sciadv.abg9204}
\end{barticle}
\endbibitem

\bibitem[\protect\citeauthoryear{Zhou et~al.}{2020}]{Zhou2020}
\begin{barticle}
\bauthor{\bsnm{Zhou}, \binits{H.}},
\bauthor{\bsnm{Choi}, \binits{J.}},
\bauthor{\bsnm{Choi}, \binits{S.}},
\bauthor{\bsnm{Landig}, \binits{R.}},
\bauthor{\bsnm{Douglas}, \binits{A.M.}},
\bauthor{\bsnm{Isoya}, \binits{J.}},
\bauthor{\bsnm{Jelezko}, \binits{F.}},
\bauthor{\bsnm{Onoda}, \binits{S.}},
\bauthor{\bsnm{Sumiya}, \binits{H.}},
\bauthor{\bsnm{Cappellaro}, \binits{P.}},
\bauthor{\bsnm{Knowles}, \binits{H.S.}},
\bauthor{\bsnm{Park}, \binits{H.}},
\bauthor{\bsnm{Lukin}, \binits{M.D.}}:
\batitle{Quantum metrology with strongly interacting spin systems}.
\bjtitle{Phys. Rev. X.}
\bvolume{10},
\bfpage{031003}
(\byear{2020})
\doiurl{10.1103/PhysRevX.10.031003}
\end{barticle}
\endbibitem

\bibitem[\protect\citeauthoryear{Choi et~al.}{2017}]{Choi2017b}
\begin{botherref}
\oauthor{\bsnm{Choi}, \binits{S.}},
\oauthor{\bsnm{Yao}, \binits{N.}},
\oauthor{\bsnm{Lukin}, \binits{M.}}:
Quantum metrology based on strongly correlated matter.
Preprint at \url{https://arxiv.org/abs/1801.00042}
(2017)
\end{botherref}
\endbibitem

\bibitem[\protect\citeauthoryear{Block et~al.}{2023}]{Block2023}
\begin{botherref}
\oauthor{\bsnm{Block}, \binits{M.}},
\oauthor{\bsnm{Ye}, \binits{B.}},
\oauthor{\bsnm{Roberts}, \binits{B.}},
\oauthor{\bsnm{Chern}, \binits{S.}},
\oauthor{\bsnm{Wu}, \binits{W.}},
\oauthor{\bsnm{Wang}, \binits{Z.}},
\oauthor{\bsnm{Pollet}, \binits{L.}},
\oauthor{\bsnm{Davis}, \binits{E.J.}},
\oauthor{\bsnm{Halperin}, \binits{B.I.}},
\oauthor{\bsnm{Yao}, \binits{N.Y.}}:
A Universal Theory of Spin Squeezing.
Preprint at \url{https://arXiv:2301.09636}
(2023)
\end{botherref}
\endbibitem

\bibitem[\protect\citeauthoryear{Reich et~al.}{1987}]{reich1987glassy}
\begin{barticle}
\bauthor{\bsnm{Reich}, \binits{D.}},
\bauthor{\bsnm{Rosenbaum}, \binits{T.}},
\bauthor{\bsnm{Aeppli}, \binits{G.}}:
\batitle{Glassy relaxation without freezing in a random dipolar-coupled ising
  magnet}.
\bjtitle{Physical Review Letters}
\bvolume{59}(\bissue{17}),
\bfpage{1969}
(\byear{1987})
\end{barticle}
\endbibitem

\bibitem[\protect\citeauthoryear{Ghosh et~al.}{2002}]{ghosh2002coherent}
\begin{barticle}
\bauthor{\bsnm{Ghosh}, \binits{S.}},
\bauthor{\bsnm{Parthasarathy}, \binits{R.}},
\bauthor{\bsnm{Rosenbaum}, \binits{T.}},
\bauthor{\bsnm{Aeppli}, \binits{G.}}:
\batitle{Coherent spin oscillations in a disordered magnet}.
\bjtitle{Science}
\bvolume{296}(\bissue{5576}),
\bfpage{2195}--\blpage{2198}
(\byear{2002})
\end{barticle}
\endbibitem

\bibitem[\protect\citeauthoryear{Rodriguez et~al.}{2010}]{rodriguez2010study}
\begin{barticle}
\bauthor{\bsnm{Rodriguez}, \binits{J.}},
\bauthor{\bsnm{Aczel}, \binits{A.A.}},
\bauthor{\bsnm{Carlo}, \binits{J.}},
\bauthor{\bsnm{Dunsiger}, \binits{S.}},
\bauthor{\bsnm{Macdougall}, \binits{G.J.}},
\bauthor{\bsnm{Russo}, \binits{P.L.}},
\bauthor{\bsnm{Savici}, \binits{A.T.}},
\bauthor{\bsnm{Uemura}, \binits{Y.J.}},
\bauthor{\bsnm{Wiebe}, \binits{C.}},
\bauthor{\bsnm{Luke}, \binits{G.M.}}:
\batitle{Study of the ground state properties of liho x y 1- x f 4 using muon
  spin relaxation}.
\bjtitle{Physical review letters}
\bvolume{105}(\bissue{10}),
\bfpage{107203}
(\byear{2010})
\end{barticle}
\endbibitem

\bibitem[\protect\citeauthoryear{Hughes et~al.}{2023}]{Hughes2023}
\begin{botherref}
\oauthor{\bsnm{Hughes}, \binits{L.B.}},
\oauthor{\bsnm{Zhang}, \binits{Z.}},
\oauthor{\bsnm{Jin}, \binits{C.}},
\oauthor{\bsnm{Meynell}, \binits{S.A.}},
\oauthor{\bsnm{Ye}, \binits{B.}},
\oauthor{\bsnm{Wu}, \binits{W.}},
\oauthor{\bsnm{Wang}, \binits{Z.}},
\oauthor{\bsnm{Davis}, \binits{E.J.}},
\oauthor{\bsnm{Mates}, \binits{T.E.}},
\oauthor{\bsnm{Yao}, \binits{N.Y.}},
\oauthor{\bsnm{Mukherjee}, \binits{K.}},
\oauthor{\bsnm{Bleszynski~Jayich}, \binits{A.C.}}:
Two-dimensional spin systems in pecvd-grown diamond with tunable density and
  long coherence for enhanced quantum sensing and simulation.
APL Mater.
\textbf{11}(2)
(2023)
\doiurl{10.1063/5.0133501}
\end{botherref}
\endbibitem

\bibitem[\protect\citeauthoryear{Davis et~al.}{2023}]{Davis2023}
\begin{barticle}
\bauthor{\bsnm{Davis}, \binits{E.J.}},
\bauthor{\bsnm{Ye}, \binits{B.}},
\bauthor{\bsnm{Machado}, \binits{F.}},
\bauthor{\bsnm{Meynell}, \binits{S.A.}},
\bauthor{\bsnm{Wu}, \binits{W.}},
\bauthor{\bsnm{Mittiga}, \binits{T.}},
\bauthor{\bsnm{Schenken}, \binits{W.}},
\bauthor{\bsnm{Joos}, \binits{M.}},
\bauthor{\bsnm{Kobrin}, \binits{B.}},
\bauthor{\bsnm{Lyu}, \binits{Y.}},
\bauthor{\bsnm{Wang}, \binits{Z.}},
\bauthor{\bsnm{Bluvstein}, \binits{D.}},
\bauthor{\bsnm{Choi}, \binits{S.}},
\bauthor{\bsnm{Zu}, \binits{C.}},
\bauthor{\bsnm{B.}, \binits{J.A.C.}},
\bauthor{\bsnm{Yao}, \binits{N.Y.}}:
\batitle{Probing many-body dynamics in a two-dimensional dipolar spin
  ensemble}.
\bjtitle{Nat. Phys.}
\bvolume{19},
\bfpage{836}--\blpage{844}
(\byear{2023})
\doiurl{10.1038/s41567-023-01944-5}
\end{barticle}
\endbibitem

\bibitem[\protect\citeauthoryear{Yao et~al.}{2014}]{Yao2014}
\begin{barticle}
\bauthor{\bsnm{Yao}, \binits{N.Y.}},
\bauthor{\bsnm{Laumann}, \binits{C.R.}},
\bauthor{\bsnm{Gopalakrishnan}, \binits{S.}},
\bauthor{\bsnm{Knap}, \binits{M.}},
\bauthor{\bsnm{Müller}, \binits{M.}},
\bauthor{\bsnm{Demler}, \binits{E.A.}},
\bauthor{\bsnm{Lukin}, \binits{M.D.}}:
\batitle{Many-body localization in dipolar systems}.
\bjtitle{Phys. Rev. Lett.}
\bvolume{113},
\bfpage{243002}
(\byear{2014})
\end{barticle}
\endbibitem

\bibitem[\protect\citeauthoryear{Burin}{2015}]{Burin2015}
\begin{barticle}
\bauthor{\bsnm{Burin}, \binits{A.L.}}:
\batitle{Many-body delocalization in a strongly disordered system with
  long-range interactions: Finite-size scaling}.
\bjtitle{Phys. Rev. B.}
\bvolume{91},
\bfpage{094202}
(\byear{2015})
\doiurl{10.1103/PhysRevB.91.094202}
\end{barticle}
\endbibitem

\bibitem[\protect\citeauthoryear{Abanin et~al.}{2019}]{Abanin2019}
\begin{barticle}
\bauthor{\bsnm{Abanin}, \binits{D.A.}},
\bauthor{\bsnm{Altman}, \binits{E.}},
\bauthor{\bsnm{Bloch}, \binits{I.}},
\bauthor{\bsnm{Serbyn}, \binits{M.}}:
\batitle{Colloquium: Many-body localization, thermalization, and entanglement}.
\bjtitle{Rev. Mod. Phys.}
\bvolume{91},
\bfpage{021001}
(\byear{2019})
\end{barticle}
\endbibitem

\bibitem[\protect\citeauthoryear{Yao et~al.}{2018}]{yao2018quantum}
\begin{barticle}
\bauthor{\bsnm{Yao}, \binits{N.Y.}},
\bauthor{\bsnm{Zaletel}, \binits{M.P.}},
\bauthor{\bsnm{Stamper-Kurn}, \binits{D.M.}},
\bauthor{\bsnm{Vishwanath}, \binits{A.}}:
\batitle{A quantum dipolar spin liquid}.
\bjtitle{Nature Physics}
\bvolume{14}(\bissue{4}),
\bfpage{405}--\blpage{410}
(\byear{2018})
\end{barticle}
\endbibitem

\bibitem[\protect\citeauthoryear{Zou et~al.}{2017}]{zou2017frustrated}
\begin{barticle}
\bauthor{\bsnm{Zou}, \binits{H.}},
\bauthor{\bsnm{Zhao}, \binits{E.}},
\bauthor{\bsnm{Liu}, \binits{W.V.}}:
\batitle{Frustrated magnetism of dipolar molecules on a square optical lattice:
  prediction of a quantum paramagnetic ground state}.
\bjtitle{Physical Review Letters}
\bvolume{119}(\bissue{5}),
\bfpage{050401}
(\byear{2017})
\end{barticle}
\endbibitem

\bibitem[\protect\citeauthoryear{Tokuda et~al.}{2007}]{Tokuda2007}
\begin{barticle}
\bauthor{\bsnm{Tokuda}, \binits{N.}},
\bauthor{\bsnm{Umezawa}, \binits{H.}},
\bauthor{\bsnm{Saito}, \binits{T.}},
\bauthor{\bsnm{Yamabe}, \binits{K.}},
\bauthor{\bsnm{Okushi}, \binits{H.}},
\bauthor{\bsnm{Yamasaki}, \binits{S.}}:
\batitle{Surface roughening of diamond (001) films during homoepitaxial growth
  in heavy boron doping}.
\bjtitle{Diam. Rel. Mater.}
\bvolume{16}(\bissue{4-7}),
\bfpage{767}--\blpage{770}
(\byear{2007})
\doiurl{10.1016/j.diamond.2006.12.024}
\end{barticle}
\endbibitem

\bibitem[\protect\citeauthoryear{de~Theije et~al.}{2000}]{deTheije2000}
\begin{barticle}
\bauthor{\bsnm{Theije}, \binits{F.K.}},
\bauthor{\bsnm{Schermer}, \binits{J.J.}},
\bauthor{\bsnm{Enckevort}, \binits{W.J.P.}}:
\batitle{Effects of nitrogen impurities on the cvd growth of diamond: step
  bunching in theory and experiment}.
\bjtitle{Diam. Rel. Mater.}
\bvolume{9}(\bissue{8}),
\bfpage{1439}--\blpage{1449}
(\byear{2000})
\doiurl{10.1016/S0925-9635(00)00261-2}
\end{barticle}
\endbibitem

\bibitem[\protect\citeauthoryear{Yamada et~al.}{2015}]{Yamada2015}
\begin{barticle}
\bauthor{\bsnm{Yamada}, \binits{H.}},
\bauthor{\bsnm{Chayahara}, \binits{A.}},
\bauthor{\bsnm{Mokuno}, \binits{Y.}}:
\batitle{Effects of intentionally introduced nitrogen and substrate temperature
  on growth of diamond bulk single crystals}.
\bjtitle{Jap. J. Appl. Phys.}
\bvolume{55},
\bfpage{01}--\blpage{07}
(\byear{2015})
\doiurl{10.7567/jjap.55.01ac07}
\end{barticle}
\endbibitem

\bibitem[\protect\citeauthoryear{Watanabe et~al.}{2009}]{Watanabe2009}
\begin{barticle}
\bauthor{\bsnm{Watanabe}, \binits{H.}},
\bauthor{\bsnm{Kitamura}, \binits{T.}},
\bauthor{\bsnm{Nakashima}, \binits{S.}},
\bauthor{\bsnm{Shikata}, \binits{S.}}:
\batitle{Cathodoluminescence characterization of a nitrogen-doped homoepitaxial
  diamond thin film}.
\bjtitle{J. Appl. Phys.}
\bvolume{105}(\bissue{9}),
\bfpage{093529}
(\byear{2009})
\doiurl{10.1063/1.3117214}
\end{barticle}
\endbibitem

\bibitem[\protect\citeauthoryear{Khomich et~al.}{2015}]{Khomich2015}
\begin{barticle}
\bauthor{\bsnm{Khomich}, \binits{A.A.}},
\bauthor{\bsnm{Kudryavtsev}, \binits{O.S.}},
\bauthor{\bsnm{Bolshakov}, \binits{A.P.}},
\bauthor{\bsnm{Khomich}, \binits{V.A.}},
\bauthor{\bsnm{Ashkinazi}, \binits{E.E.}},
\bauthor{\bsnm{Ralchenko}, \binits{V.G.}},
\bauthor{\bsnm{Vlasov}, \binits{I.I.}},
\bauthor{\bsnm{Konov}, \binits{V.I.}}:
\batitle{Use of optical spectroscopy methods to determine the solubility limit
  for nitrogen in diamond single crystals synthesized by chemical vapor
  deposition}.
\bjtitle{J. Appl. Spec.}
\bvolume{82}(\bissue{2}),
\bfpage{242}--\blpage{247}
(\byear{2015})
\doiurl{10.1007/s10812-015-0092-1}
\end{barticle}
\endbibitem

\bibitem[\protect\citeauthoryear{Eichhorn et~al.}{2019}]{Eichhorn2019}
\begin{barticle}
\bauthor{\bsnm{Eichhorn}, \binits{T.}},
\bauthor{\bsnm{McLellan}, \binits{C.}},
\bauthor{\bsnm{Jayich}, \binits{A.C.B.}}:
\batitle{Optimizing the formation of depth-confined nitrogen vacancy center
  spin ensembles in diamond for quantum sensing}.
\bjtitle{Phys. Rev. Mater.}
\bvolume{3},
\bfpage{113802}
(\byear{2019})
\doiurl{10.1103/PhysRevMaterials.3.113802}
\end{barticle}
\endbibitem

\bibitem[\protect\citeauthoryear{Tallaire et~al.}{2020}]{Tallaire2020}
\begin{barticle}
\bauthor{\bsnm{Tallaire}, \binits{A.}},
\bauthor{\bsnm{Brinza}, \binits{O.}},
\bauthor{\bsnm{Huillery}, \binits{P.}},
\bauthor{\bsnm{Delord}, \binits{T.}},
\bauthor{\bsnm{Pellet-Mary}, \binits{C.}},
\bauthor{\bsnm{Staacke}, \binits{R.}}:
\batitle{High nv density in a pink cvd diamond grown with n2o addition}.
\bjtitle{Carbon}
\bvolume{170},
\bfpage{421}--\blpage{429}
(\byear{2020})
\doiurl{10.1016/j.carbon.2020.08.048}
\end{barticle}
\endbibitem

\bibitem[\protect\citeauthoryear{Samlenski et~al.}{1995}]{Samlenski1995}
\begin{barticle}
\bauthor{\bsnm{Samlenski}, \binits{R.}},
\bauthor{\bsnm{Haug}, \binits{C.}},
\bauthor{\bsnm{Brenn}, \binits{R.}},
\bauthor{\bsnm{Wild}, \binits{C.}},
\bauthor{\bsnm{Locher}, \binits{R.}},
\bauthor{\bsnm{Koidl}, \binits{P.}}:
\batitle{Incorporation of nitrogen in chemical vapor deposition diamond}.
\bjtitle{Appl. Phys. Lett.}
\bvolume{67},
\bfpage{2798}
(\byear{1995})
\doiurl{10.1063/1.114788}
\end{barticle}
\endbibitem

\bibitem[\protect\citeauthoryear{Lesik et~al.}{2014}]{Lesik2014}
\begin{barticle}
\bauthor{\bsnm{Lesik}, \binits{M.}},
\bauthor{\bsnm{Tetienne}, \binits{J.P.}},
\bauthor{\bsnm{Tallaire}, \binits{A.}},
\bauthor{\bsnm{Achard}, \binits{J.}},
\bauthor{\bsnm{Mille}, \binits{V.}},
\bauthor{\bsnm{Gicquel}, \binits{A.}},
\bauthor{\bsnm{Roch}, \binits{J.F.}},
\bauthor{\bsnm{Jacques}, \binits{V.}}:
\batitle{Perfect preferential orientation of nitrogen-vacancy defects in a
  synthetic diamond sample}.
\bjtitle{Appl. Phys. Lett.}
\bvolume{104}(\bissue{11}),
\bfpage{113107}
(\byear{2014})
\doiurl{10.1063/1.4869103}
\end{barticle}
\endbibitem

\bibitem[\protect\citeauthoryear{Michl et~al.}{2014}]{Michl2014}
\begin{barticle}
\bauthor{\bsnm{Michl}, \binits{J.}},
\bauthor{\bsnm{Teraji}, \binits{T.}},
\bauthor{\bsnm{Zaiser}, \binits{S.}},
\bauthor{\bsnm{Jakobi}, \binits{I.}},
\bauthor{\bsnm{Waldherr}, \binits{G.}},
\bauthor{\bsnm{Dolde}, \binits{F.}},
\bauthor{\bsnm{Neumann}, \binits{P.}},
\bauthor{\bsnm{Doherty}, \binits{M.W.}},
\bauthor{\bsnm{Manson}, \binits{N.B.}},
\bauthor{\bsnm{Isoya}, \binits{J.}},
\bauthor{\bsnm{Wrachtrup}, \binits{J.}}:
\batitle{Perfect alignment and preferential orientation of nitrogen-vacancy
  centers during chemical vapor deposition diamond growth on (111) surfaces}.
\bjtitle{Appl. Phys. Lett.}
\bvolume{104}(\bissue{10}),
\bfpage{102407}
(\byear{2014})
\doiurl{10.1063/1.4868128}
\end{barticle}
\endbibitem

\bibitem[\protect\citeauthoryear{Fukui et~al.}{2014}]{Fukui2014}
\begin{barticle}
\bauthor{\bsnm{Fukui}, \binits{T.}},
\bauthor{\bsnm{Doi}, \binits{Y.}},
\bauthor{\bsnm{Miyazaki}, \binits{T.}},
\bauthor{\bsnm{Miyamoto}, \binits{Y.}},
\bauthor{\bsnm{Kato}, \binits{H.}},
\bauthor{\bsnm{Matsumoto}, \binits{T.}},
\bauthor{\bsnm{Makino}, \binits{T.}},
\bauthor{\bsnm{Yamasaki}, \binits{S.}},
\bauthor{\bsnm{Morimoto}, \binits{R.}},
\bauthor{\bsnm{Tokuda}, \binits{N.}},
\bauthor{\bsnm{Hatano}, \binits{M.}},
\bauthor{\bsnm{Sakagawa}, \binits{Y.}},
\bauthor{\bsnm{Morishita}, \binits{H.}},
\bauthor{\bsnm{Tashima}, \binits{T.}},
\bauthor{\bsnm{Miwa}, \binits{S.}},
\bauthor{\bsnm{Suzuki}, \binits{Y.}},
\bauthor{\bsnm{Mizuochi}, \binits{N.}}:
\batitle{Perfect selective alignment of nitrogen-vacancy centers in diamond}.
\bjtitle{Appl. Phys. Exp.}
\bvolume{7}(\bissue{5}),
\bfpage{055201}
(\byear{2014})
\doiurl{10.7567/APEX.7.055201}
\end{barticle}
\endbibitem

\bibitem[\protect\citeauthoryear{Miyazaki et~al.}{2014}]{Miyazaki2014}
\begin{barticle}
\bauthor{\bsnm{Miyazaki}, \binits{T.}},
\bauthor{\bsnm{Miyamoto}, \binits{Y.}},
\bauthor{\bsnm{Makino}, \binits{T.}},
\bauthor{\bsnm{Kato}, \binits{H.}},
\bauthor{\bsnm{Yamasaki}, \binits{S.}},
\bauthor{\bsnm{Fukui}, \binits{Y.} \bsuffix{T.~andDoi}},
\bauthor{\bsnm{Tokuda}, \binits{N.}},
\bauthor{\bsnm{Hatano}, \binits{M.}},
\bauthor{\bsnm{Mizuochi}, \binits{N.}}:
\batitle{Atomistic mechanism of perfect alignment of nitrogen-vacancy centers
  in diamond}.
\bjtitle{Appl. Phys. Lett.}
\bvolume{105}(\bissue{26}),
\bfpage{261601}
(\byear{2014})
\doiurl{10.1063/1.4904988}
\end{barticle}
\endbibitem

\bibitem[\protect\citeauthoryear{Ozawa et~al.}{2017}]{Ozawa2017}
\begin{barticle}
\bauthor{\bsnm{Ozawa}, \binits{H.}},
\bauthor{\bsnm{Tahara}, \binits{K.}},
\bauthor{\bsnm{Ishiwata}, \binits{H.}},
\bauthor{\bsnm{Hatano}, \binits{M.}},
\bauthor{\bsnm{Iwasaki}, \binits{T.}}:
\batitle{Formation of perfectly aligned nitrogen-vacancy-center ensembles in
  chemical-vapor-deposition-grown diamond (111)}.
\bjtitle{Appl. Phys. Exp.}
\bvolume{10}(\bissue{4}),
\bfpage{045501}
(\byear{2017})
\doiurl{10.7567/APEX.10.045501}
\end{barticle}
\endbibitem

\bibitem[\protect\citeauthoryear{Osterkamp et~al.}{2019}]{Osterkamp2019}
\begin{barticle}
\bauthor{\bsnm{Osterkamp}, \binits{C.}},
\bauthor{\bsnm{Mangold}, \binits{M.}},
\bauthor{\bsnm{Lang}, \binits{J.}},
\bauthor{\bsnm{Balasubramanian}, \binits{P.}},
\bauthor{\bsnm{Teraji}, \binits{T.}},
\bauthor{\bsnm{Naydenov}, \binits{B.}},
\bauthor{\bsnm{Jelezko}, \binits{F.}}:
\batitle{Engineering preferentially-aligned nitrogen-vacancy centre ensembles
  in cvd grown diamond}.
\bjtitle{Sci. Rep.}
\bvolume{9}(\bissue{1}),
\bfpage{1}--\blpage{7}
(\byear{2019})
\doiurl{10.1038/s41598-019-42314-7}
\end{barticle}
\endbibitem

\bibitem[\protect\citeauthoryear{Sakaguchi et~al.}{1999}]{Sakaguchi1999}
\begin{barticle}
\bauthor{\bsnm{Sakaguchi}, \binits{I.}},
\bauthor{\bsnm{Nishitani-Gamo}, \binits{M.}},
\bauthor{\bsnm{Loh}, \binits{K.P.}},
\bauthor{\bsnm{Hajime}, \binits{H.}},
\bauthor{\bsnm{Toshihiro}, \binits{A.}}:
\batitle{Homoepitaxial growth and hydrogen incorporation on the chemical vapor
  deposited (111) diamond}.
\bjtitle{J. Appl. Phys.}
\bvolume{86}(\bissue{3}),
\bfpage{1306}--\blpage{1310}
(\byear{1999})
\doiurl{10.1063/1.370886}
\end{barticle}
\endbibitem

\bibitem[\protect\citeauthoryear{Kasu et~al.}{2003}]{Kasu2003}
\begin{barticle}
\bauthor{\bsnm{Kasu}, \binits{M.}},
\bauthor{\bsnm{Makimoto}, \binits{T.}},
\bauthor{\bsnm{Ebert}, \binits{W.}},
\bauthor{\bsnm{Kohn}, \binits{E.}}:
\batitle{Formation of stacking faults containing microtwins in (111)
  chemical-vapor-deposited diamond homoepitaxial layers}.
\bjtitle{Appl. Phys. Lett.}
\bvolume{83}(\bissue{17}),
\bfpage{3465}--\blpage{3467}
(\byear{2003})
\doiurl{10.1063/1.1622105}
\end{barticle}
\endbibitem

\bibitem[\protect\citeauthoryear{Tokuda et~al.}{2008}]{Tokuda2008}
\begin{barticle}
\bauthor{\bsnm{Tokuda}, \binits{N.}},
\bauthor{\bsnm{Umezawa}, \binits{H.}},
\bauthor{\bsnm{Ri}, \binits{S.G.}},
\bauthor{\bsnm{Ogura}, \binits{M.}},
\bauthor{\bsnm{Yamabe}, \binits{K.}},
\bauthor{\bsnm{Okushi}, \binits{H.}},
\bauthor{\bsnm{Yamasaki}, \binits{S.}}:
\batitle{Atomically flat diamond (111) surface formation by homoepitaxial
  lateral growth}.
\bjtitle{Diam. Rel. Mater.}
\bvolume{17}(\bissue{7-10}),
\bfpage{1051}--\blpage{1054}
(\byear{2008})
\doiurl{10.1016/j.diamond.2008.01.089}
\end{barticle}
\endbibitem

\bibitem[\protect\citeauthoryear{K.}{2015}]{Larsson2015}
\begin{bchapter}
\bauthor{\bsnm{K.}, \binits{L.}}:
\bctitle{Surface chemistry of dimaond}.
In: \beditor{\bsnm{Yang}, \binits{N.}} (ed.)
\bbtitle{Novel Aspects of Diamond From Growth to Applications},
pp. \bfpage{53}--\blpage{83}.
\bpublisher{Springer},
\blocation{Switzerland}
(\byear{2015}).
\doiurl{10.1007/978-3-319-09834-0_3}
\end{bchapter}
\endbibitem

\bibitem[\protect\citeauthoryear{Meynell et~al.}{2020}]{Meynell2020}
\begin{barticle}
\bauthor{\bsnm{Meynell}, \binits{S.}},
\bauthor{\bsnm{McLellan}, \binits{C.}},
\bauthor{\bsnm{Hughes}, \binits{L.B.}},
\bauthor{\bsnm{Wang}, \binits{W.}},
\bauthor{\bsnm{Mates}, \binits{T.E.}},
\bauthor{\bsnm{Mukherjee}, \binits{K.}},
\bauthor{\bsnm{Jayich}, \binits{A.C.B.}}:
\batitle{Engineering quantum-coherent defects: the role of substrate miscut in
  chemical vapor deposition diamond growth}.
\bjtitle{Appl. Phys. Lett.}
\bvolume{117},
\bfpage{194001}
(\byear{2020})
\end{barticle}
\endbibitem

\bibitem[\protect\citeauthoryear{Kato et~al.}{2007}]{Kato2007}
\begin{barticle}
\bauthor{\bsnm{Kato}, \binits{H.}},
\bauthor{\bsnm{Makino}, \binits{T.}},
\bauthor{\bsnm{Yamasaki}, \binits{S.}},
\bauthor{\bsnm{Okushi}, \binits{H.}}:
\batitle{N-type diamond growth by phosphorus doping on (0 0 1)-oriented
  surface}.
\bjtitle{J. Phys. D. Appl. Phys.}
\bvolume{40}(\bissue{20}),
\bfpage{6189}--\blpage{6200}
(\byear{2007})
\doiurl{10.1088/0022-3727/40/20/S05}
\end{barticle}
\endbibitem

\bibitem[\protect\citeauthoryear{Mortet et~al.}{2022}]{Mortet2022}
\begin{barticle}
\bauthor{\bsnm{Mortet}, \binits{V.}},
\bauthor{\bsnm{Taylor}, \binits{A.}},
\bauthor{\bsnm{Davydova}, \binits{M.}},
\bauthor{\bsnm{Jiránek}, \binits{J.}},
\bauthor{\bsnm{Fekete}, \binits{L.}},
\bauthor{\bsnm{Klimša}, \binits{L.}},
\bauthor{\bsnm{Šimek}, \binits{D.}},
\bauthor{\bsnm{Lambert}, \binits{N.}},
\bauthor{\bsnm{Sedláková}, \binits{S.}},
\bauthor{\bsnm{Kopeček}, \binits{J.}},
\bauthor{\bsnm{Hazdra}, \binits{P.}}:
\batitle{Effect of substrate crystalline orientation on boron-doped
  homoepitaxial diamond growth}.
\bjtitle{Diamond and Related Materials}
\bvolume{122},
\bfpage{108887}
(\byear{2022})
\doiurl{10.1016/j.diamond.2022.108887}
\end{barticle}
\endbibitem

\bibitem[\protect\citeauthoryear{Nakano et~al.}{2022}]{Nakano2022}
\begin{barticle}
\bauthor{\bsnm{Nakano}, \binits{Y.}},
\bauthor{\bsnm{Zhang}, \binits{X.}},
\bauthor{\bsnm{Kobayashi}, \binits{K.}},
\bauthor{\bsnm{Matsumoto}, \binits{T.}},
\bauthor{\bsnm{Inokuma}, \binits{T.}},
\bauthor{\bsnm{Yamasaki}, \binits{S.}},
\bauthor{\bsnm{Nebel}, \binits{C.E.}},
\bauthor{\bsnm{Tokuda}, \binits{N.}}:
\batitle{Impact of nitrogen doping on homoepitaxial diamond (111) growth}.
\bjtitle{Diam. Rel. Mater.}
\bvolume{125},
\bfpage{108997}
(\byear{2022})
\doiurl{10.1016/j.diamond.2022.108997}
\end{barticle}
\endbibitem

\bibitem[\protect\citeauthoryear{Tallaire et~al.}{2015}]{Tallaire2015}
\begin{barticle}
\bauthor{\bsnm{Tallaire}, \binits{A.}},
\bauthor{\bsnm{Lesik}, \binits{M.}},
\bauthor{\bsnm{Jacques}, \binits{V.}},
\bauthor{\bsnm{Pezzagna}, \binits{S.}},
\bauthor{\bsnm{Mille}, \binits{V.}},
\bauthor{\bsnm{Brinza}, \binits{O.}},
\bauthor{\bsnm{Meijer}, \binits{J.}},
\bauthor{\bsnm{Abel}, \binits{B.}},
\bauthor{\bsnm{Roch}, \binits{J.F.}},
\bauthor{\bsnm{Gicquel}, \binits{A.}},
\bauthor{\bsnm{J.Achard}}:
\batitle{Temperature dependent creation of nitrogen-vacancy centers in single
  crystal cvd diamond layers}.
\bjtitle{Diam. Rel. Mater.}
\bvolume{51},
\bfpage{55}--\blpage{60}
(\byear{2015})
\end{barticle}
\endbibitem

\bibitem[\protect\citeauthoryear{Ishiwata et~al.}{2017}]{Ishiwata2017}
\begin{barticle}
\bauthor{\bsnm{Ishiwata}, \binits{H.}},
\bauthor{\bsnm{Nakajima}, \binits{M.}},
\bauthor{\bsnm{Tahara}, \binits{K.}},
\bauthor{\bsnm{Ozawa}, \binits{H.}},
\bauthor{\bsnm{Iwasaki}, \binits{T.}},
\bauthor{\bsnm{Hatano}, \binits{M.}}:
\batitle{Perfectly aligned shallow ensemble nitrogen-vacancy centers in (111)
  diamond}.
\bjtitle{Appl. Phys. Lett.}
\bvolume{111},
\bfpage{043103}
(\byear{2017})
\doiurl{10.1063/1.4993160}
\end{barticle}
\endbibitem

\bibitem[\protect\citeauthoryear{Wolf et~al.}{2015}]{Wolf2015}
\begin{barticle}
\bauthor{\bsnm{Wolf}, \binits{T.}},
\bauthor{\bsnm{Neumann}, \binits{P.}},
\bauthor{\bsnm{Nakamura}, \binits{K.}},
\bauthor{\bsnm{Sumiya}, \binits{H.}},
\bauthor{\bsnm{Ohshima}, \binits{T.}},
\bauthor{\bsnm{Isoya}, \binits{J.}},
\bauthor{\bsnm{Wrachtrup}, \binits{J.}}:
\batitle{Subpicotesla diamond magnetometry}.
\bjtitle{Phys. Rev. X}
\bvolume{5},
\bfpage{041001}
(\byear{2015})
\doiurl{10.1103/PhysRevX.5.041001}
\end{barticle}
\endbibitem

\bibitem[\protect\citeauthoryear{Osterkamp et~al.}{2020}]{Osterkamp2020}
\begin{barticle}
\bauthor{\bsnm{Osterkamp}, \binits{C.}},
\bauthor{\bsnm{Balasubramanian}, \binits{P.}},
\bauthor{\bsnm{Wolff}, \binits{G.}},
\bauthor{\bsnm{Teraji}, \binits{T.}},
\bauthor{\bsnm{Nesladek}, \binits{M.}},
\bauthor{\bsnm{Jelezko}, \binits{F.}}:
\batitle{Benchmark for synthesized diamond sensors based on isotopically
  engineered nitrogen-vacancy spin ensembles for magnetometry applications}.
\bjtitle{Adv. Quantum Technol.}
\bvolume{3},
\bfpage{2000074}
(\byear{2020})
\doiurl{10.1002/qute.202000074}
\end{barticle}
\endbibitem

\bibitem[\protect\citeauthoryear{Perlin et~al.}{2020}]{Perlin2020}
\begin{barticle}
\bauthor{\bsnm{Perlin}, \binits{M.A.}},
\bauthor{\bsnm{Qu}, \binits{C.}},
\bauthor{\bsnm{Rey}, \binits{A.M.}}:
\batitle{Spin squeezing with short-range spin-exchange interactions}.
\bjtitle{Phys. Rev. Lett.}
\bvolume{125},
\bfpage{223401}
(\byear{2020})
\doiurl{10.1103/PhysRevLett.125.223401}
\end{barticle}
\endbibitem

\bibitem[\protect\citeauthoryear{Bornet et~al.}{2023}]{Bornet2023}
\begin{barticle}
\bauthor{\bsnm{Bornet}, \binits{G.}},
\bauthor{\bsnm{Emperauger}, \binits{G.}},
\bauthor{\bsnm{Chen}, \binits{C.}},
\bauthor{\bsnm{Ye}, \binits{B.}},
\bauthor{\bsnm{Block}, \binits{M.}},
\bauthor{\bsnm{Bintz}, \binits{M.}},
\bauthor{\bsnm{Boyd}, \binits{J.A.}},
\bauthor{\bsnm{Barredo}, \binits{D.}},
\bauthor{\bsnm{Comparin}, \binits{T.}},
\bauthor{\bsnm{Mezzacapo}, \binits{F.}},
\bauthor{\bsnm{Roscilde}, \binits{T.}},
\bauthor{\bsnm{Lahaye}, \binits{T.}},
\bauthor{\bsnm{Yao}, \binits{N.Y.}},
\bauthor{\bsnm{Browaeys}, \binits{A.}}:
\batitle{Scalable spin squeezing in a dipolar rydberg atom array}.
\bjtitle{Nature}
\bvolume{621},
\bfpage{728}--\blpage{733}
(\byear{2023})
\doiurl{10.1038/s41586-023-06414-9}
\end{barticle}
\endbibitem

\bibitem[\protect\citeauthoryear{Dréau et~al.}{2011}]{Dreau2011}
\begin{barticle}
\bauthor{\bsnm{Dréau}, \binits{A.}},
\bauthor{\bsnm{Lesik}, \binits{M.}},
\bauthor{\bsnm{Rondin}, \binits{L.}},
\bauthor{\bsnm{Spinicelli}, \binits{P.}},
\bauthor{\bsnm{Arcizet}, \binits{O.}},
\bauthor{\bsnm{Roch}, \binits{J.-F.}},
\bauthor{\bsnm{Jacques}, \binits{V.}}:
\batitle{Avoiding power broadening in optically detected magnetic resonance of
  single nv defects for enhanced dc magnetic field sensitivity}.
\bjtitle{Phys. Rev. B}
\bvolume{84},
\bfpage{195204}
(\byear{2011})
\doiurl{10.1103/PhysRevB.84.195204}
\end{barticle}
\endbibitem

\bibitem[\protect\citeauthoryear{Chou et~al.}{2017}]{Chou2017}
\begin{barticle}
\bauthor{\bsnm{Chou}, \binits{J.P.}},
\bauthor{\bsnm{Retzker}, \binits{A.}},
\bauthor{\bsnm{Gali}, \binits{A.}}:
\batitle{Nitrogen-terminated diamond (111) surface for room-temperature quantum
  sensing and simulation}.
\bjtitle{Nano. Lett.}
\bvolume{17}(\bissue{4}),
\bfpage{2294}--\blpage{2298}
(\byear{2017})
\doiurl{10.1021/acs.nanolett.6b05023}
\end{barticle}
\endbibitem

\bibitem[\protect\citeauthoryear{Chou and Gali}{2017}]{Gali2017}
\begin{barticle}
\bauthor{\bsnm{Chou}, \binits{J.P.}},
\bauthor{\bsnm{Gali}, \binits{A.}}:
\batitle{Nitrogen-vacancy diamond sensor: Novel diamond surfaces from ab initio
  simulations}.
\bjtitle{MRS Comm.}
\bvolume{7}(\bissue{3}),
\bfpage{551}--\blpage{552}
(\byear{2017})
\doiurl{10.1557/mrc.2017.75}
\end{barticle}
\endbibitem

\bibitem[\protect\citeauthoryear{Parks et~al.}{2018}]{Parks2018}
\begin{barticle}
\bauthor{\bsnm{Parks}, \binits{S.M.}},
\bauthor{\bsnm{Grotea}, \binits{R.R.}},
\bauthor{\bsnm{Hopper}, \binits{D.A.}},
\bauthor{\bsnm{Bassett}, \binits{L.C.}}:
\batitle{Fabrication of (111)-faced single-crystal diamond plates by laser
  nucleated cleaving}.
\bjtitle{Diam. Rel. Mater.}
\bvolume{84},
\bfpage{20}--\blpage{25}
(\byear{2018})
\doiurl{10.1016/j.diamond.2018.02.013}
\end{barticle}
\endbibitem

\bibitem[\protect\citeauthoryear{McLellan et~al.}{2016}]{McLellan2016}
\begin{barticle}
\bauthor{\bsnm{McLellan}, \binits{C.A.}},
\bauthor{\bsnm{Myers}, \binits{B.A.}},
\bauthor{\bsnm{Kraemer}, \binits{S.}},
\bauthor{\bsnm{Ohno}, \binits{K.}},
\bauthor{\bsnm{Awschalom}, \binits{D.D.}},
\bauthor{\bsnm{Jayich}, \binits{A.C.B.}}:
\batitle{Patterned formation of highly coherent nitrogen-vacancy centers using
  a focused electron irradiation technique}.
\bjtitle{Nano. Lett.}
\bvolume{16},
\bfpage{2450}--\blpage{2454}
(\byear{2016})
\doiurl{10.1021/acs.nanolett.5b05304}
\end{barticle}
\endbibitem

\end{thebibliography}



\begin{thebibliography}{12}
\ifx \bisbn   \undefined \def \bisbn  #1{ISBN #1}\fi
\ifx \binits  \undefined \def \binits#1{#1}\fi
\ifx \bauthor  \undefined \def \bauthor#1{#1}\fi
\ifx \batitle  \undefined \def \batitle#1{#1}\fi
\ifx \bjtitle  \undefined \def \bjtitle#1{#1}\fi
\ifx \bvolume  \undefined \def \bvolume#1{\textbf{#1}}\fi
\ifx \byear  \undefined \def \byear#1{#1}\fi
\ifx \bissue  \undefined \def \bissue#1{#1}\fi
\ifx \bfpage  \undefined \def \bfpage#1{#1}\fi
\ifx \blpage  \undefined \def \blpage #1{#1}\fi
\ifx \burl  \undefined \def \burl#1{\textsf{#1}}\fi
\ifx \doiurl  \undefined \def \doiurl#1{\url{https://doi.org/#1}}\fi
\ifx \betal  \undefined \def \betal{\textit{et al.}}\fi
\ifx \binstitute  \undefined \def \binstitute#1{#1}\fi
\ifx \binstitutionaled  \undefined \def \binstitutionaled#1{#1}\fi
\ifx \bctitle  \undefined \def \bctitle#1{#1}\fi
\ifx \beditor  \undefined \def \beditor#1{#1}\fi
\ifx \bpublisher  \undefined \def \bpublisher#1{#1}\fi
\ifx \bbtitle  \undefined \def \bbtitle#1{#1}\fi
\ifx \bedition  \undefined \def \bedition#1{#1}\fi
\ifx \bseriesno  \undefined \def \bseriesno#1{#1}\fi
\ifx \blocation  \undefined \def \blocation#1{#1}\fi
\ifx \bsertitle  \undefined \def \bsertitle#1{#1}\fi
\ifx \bsnm \undefined \def \bsnm#1{#1}\fi
\ifx \bsuffix \undefined \def \bsuffix#1{#1}\fi
\ifx \bparticle \undefined \def \bparticle#1{#1}\fi
\ifx \barticle \undefined \def \barticle#1{#1}\fi
\bibcommenthead
\ifx \bconfdate \undefined \def \bconfdate #1{#1}\fi
\ifx \botherref \undefined \def \botherref #1{#1}\fi
\ifx \url \undefined \def \url#1{\textsf{#1}}\fi
\ifx \bchapter \undefined \def \bchapter#1{#1}\fi
\ifx \bbook \undefined \def \bbook#1{#1}\fi
\ifx \bcomment \undefined \def \bcomment#1{#1}\fi
\ifx \oauthor \undefined \def \oauthor#1{#1}\fi
\ifx \citeauthoryear \undefined \def \citeauthoryear#1{#1}\fi
\ifx \endbibitem  \undefined \def \endbibitem {}\fi
\ifx \bconflocation  \undefined \def \bconflocation#1{#1}\fi
\ifx \arxivurl  \undefined \def \arxivurl#1{\textsf{#1}}\fi
\csname PreBibitemsHook\endcsname

\bibitem[\protect\citeauthoryear{Meynell}{2024}]{Meynell2023}
\begin{botherref}
\oauthor{\bsnm{Meynell}, \binits{S.A.}}:
Dimensionality in dipolar quantum systems.
PhD thesis,
University of California Santa Barbara
(2024)
\end{botherref}
\endbibitem

\bibitem[\protect\citeauthoryear{Cao et~al.}{1996}]{Cao1996}
\begin{barticle}
\bauthor{\bsnm{Cao}, \binits{G.Z.}},
\bauthor{\bsnm{Schermer}, \binits{J.J.}},
\bauthor{\bsnm{Enckevort}, \binits{W.J.P.}},
\bauthor{\bsnm{Elst}, \binits{L.J.} \bsuffix{W.A.L.M. abd~Giling}}:
\batitle{Growth of {100} textured diamond films by the addition of nitrogen}.
\bjtitle{J. Appl. Phys.}
\bvolume{79},
\bfpage{1357}--\blpage{1364}
(\byear{1996})
\doiurl{10.1063/1.361033}
\end{barticle}
\endbibitem

\bibitem[\protect\citeauthoryear{Tallaire et~al.}{2015}]{Tallaire2015}
\begin{barticle}
\bauthor{\bsnm{Tallaire}, \binits{A.}},
\bauthor{\bsnm{Lesik}, \binits{M.}},
\bauthor{\bsnm{Jacques}, \binits{V.}},
\bauthor{\bsnm{Pezzagna}, \binits{S.}},
\bauthor{\bsnm{Mille}, \binits{V.}},
\bauthor{\bsnm{Brinza}, \binits{O.}},
\bauthor{\bsnm{Meijer}, \binits{J.}},
\bauthor{\bsnm{Abel}, \binits{B.}},
\bauthor{\bsnm{Roch}, \binits{J.F.}},
\bauthor{\bsnm{Gicquel}, \binits{A.}},
\bauthor{\bsnm{J.Achard}}:
\batitle{Temperature dependent creation of nitrogen-vacancy centers in single
  crystal cvd diamond layers}.
\bjtitle{Diam. Rel. Mater.}
\bvolume{51},
\bfpage{55}--\blpage{60}
(\byear{2015})
\end{barticle}
\endbibitem

\bibitem[\protect\citeauthoryear{Kato et~al.}{2007}]{Kato2007}
\begin{barticle}
\bauthor{\bsnm{Kato}, \binits{H.}},
\bauthor{\bsnm{Makino}, \binits{T.}},
\bauthor{\bsnm{Yamasaki}, \binits{S.}},
\bauthor{\bsnm{Okushi}, \binits{H.}}:
\batitle{N-type diamond growth by phosphorus doping on (0 0 1)-oriented
  surface}.
\bjtitle{J. Phys. D. Appl. Phys.}
\bvolume{40}(\bissue{20}),
\bfpage{6189}--\blpage{6200}
(\byear{2007})
\doiurl{10.1088/0022-3727/40/20/S05}
\end{barticle}
\endbibitem

\bibitem[\protect\citeauthoryear{Ozawa et~al.}{2017}]{Ozawa2017}
\begin{barticle}
\bauthor{\bsnm{Ozawa}, \binits{H.}},
\bauthor{\bsnm{Tahara}, \binits{K.}},
\bauthor{\bsnm{Ishiwata}, \binits{H.}},
\bauthor{\bsnm{Hatano}, \binits{M.}},
\bauthor{\bsnm{Iwasaki}, \binits{T.}}:
\batitle{Formation of perfectly aligned nitrogen-vacancy-center ensembles in
  chemical-vapor-deposition-grown diamond (111)}.
\bjtitle{Appl. Phys. Exp.}
\bvolume{10}(\bissue{4}),
\bfpage{045501}
(\byear{2017})
\doiurl{10.7567/APEX.10.045501}
\end{barticle}
\endbibitem

\bibitem[\protect\citeauthoryear{Meynell et~al.}{2020}]{Meynell2020}
\begin{barticle}
\bauthor{\bsnm{Meynell}, \binits{S.}},
\bauthor{\bsnm{McLellan}, \binits{C.}},
\bauthor{\bsnm{Hughes}, \binits{L.B.}},
\bauthor{\bsnm{Wang}, \binits{W.}},
\bauthor{\bsnm{Mates}, \binits{T.E.}},
\bauthor{\bsnm{Mukherjee}, \binits{K.}},
\bauthor{\bsnm{Jayich}, \binits{A.C.B.}}:
\batitle{Engineering quantum-coherent defects: the role of substrate miscut in
  chemical vapor deposition diamond growth}.
\bjtitle{Appl. Phys. Lett.}
\bvolume{117},
\bfpage{194001}
(\byear{2020})
\end{barticle}
\endbibitem

\bibitem[\protect\citeauthoryear{Davis et~al.}{2023}]{Davis2023}
\begin{barticle}
\bauthor{\bsnm{Davis}, \binits{E.J.}},
\bauthor{\bsnm{Ye}, \binits{B.}},
\bauthor{\bsnm{Machado}, \binits{F.}},
\bauthor{\bsnm{Meynell}, \binits{S.A.}},
\bauthor{\bsnm{Wu}, \binits{W.}},
\bauthor{\bsnm{Mittiga}, \binits{T.}},
\bauthor{\bsnm{Schenken}, \binits{W.}},
\bauthor{\bsnm{Joos}, \binits{M.}},
\bauthor{\bsnm{Kobrin}, \binits{B.}},
\bauthor{\bsnm{Lyu}, \binits{Y.}},
\bauthor{\bsnm{Wang}, \binits{Z.}},
\bauthor{\bsnm{Bluvstein}, \binits{D.}},
\bauthor{\bsnm{Choi}, \binits{S.}},
\bauthor{\bsnm{Zu}, \binits{C.}},
\bauthor{\bsnm{B.}, \binits{J.A.C.}},
\bauthor{\bsnm{Yao}, \binits{N.Y.}}:
\batitle{Probing many-body dynamics in a two-dimensional dipolar spin
  ensemble}.
\bjtitle{Nat. Phys.}
\bvolume{19},
\bfpage{836}--\blpage{844}
(\byear{2023})
\doiurl{10.1038/s41567-023-01944-5}
\end{barticle}
\endbibitem

\bibitem[\protect\citeauthoryear{Hughes et~al.}{2023}]{Hughes2023}
\begin{botherref}
\oauthor{\bsnm{Hughes}, \binits{L.B.}},
\oauthor{\bsnm{Zhang}, \binits{Z.}},
\oauthor{\bsnm{Jin}, \binits{C.}},
\oauthor{\bsnm{Meynell}, \binits{S.A.}},
\oauthor{\bsnm{Ye}, \binits{B.}},
\oauthor{\bsnm{Wu}, \binits{W.}},
\oauthor{\bsnm{Wang}, \binits{Z.}},
\oauthor{\bsnm{Davis}, \binits{E.J.}},
\oauthor{\bsnm{Mates}, \binits{T.E.}},
\oauthor{\bsnm{Yao}, \binits{N.Y.}},
\oauthor{\bsnm{Mukherjee}, \binits{K.}},
\oauthor{\bsnm{Bleszynski~Jayich}, \binits{A.C.}}:
Two-dimensional spin systems in pecvd-grown diamond with tunable density and
  long coherence for enhanced quantum sensing and simulation.
APL Mater.
\textbf{11}(2)
(2023)
\doiurl{10.1063/5.0133501}
\end{botherref}
\endbibitem

\bibitem[\protect\citeauthoryear{Barry et~al.}{2020}]{Barry2020}
\begin{barticle}
\bauthor{\bsnm{Barry}, \binits{J.F.}},
\bauthor{\bsnm{Schloss}, \binits{J.M.}},
\bauthor{\bsnm{Bauch}, \binits{E.}},
\bauthor{\bsnm{Turner}, \binits{M.J.}},
\bauthor{\bsnm{Hart}, \binits{C.A.}},
\bauthor{\bsnm{Pham}, \binits{L.M.}},
\bauthor{\bsnm{Walsworth}, \binits{R.L.}}:
\batitle{Sensitivity optimization for nv-diamond magnetometry}.
\bjtitle{Rev. Mod. Phys.}
\bvolume{92},
\bfpage{015004}
(\byear{2020})
\doiurl{10.1103/RevModPhys.92.015004}
\end{barticle}
\endbibitem

\bibitem[\protect\citeauthoryear{Zhou et~al.}{2020}]{Zhou2020}
\begin{barticle}
\bauthor{\bsnm{Zhou}, \binits{H.}},
\bauthor{\bsnm{Choi}, \binits{J.}},
\bauthor{\bsnm{Choi}, \binits{S.}},
\bauthor{\bsnm{Landig}, \binits{R.}},
\bauthor{\bsnm{Douglas}, \binits{A.M.}},
\bauthor{\bsnm{Isoya}, \binits{J.}},
\bauthor{\bsnm{Jelezko}, \binits{F.}},
\bauthor{\bsnm{Onoda}, \binits{S.}},
\bauthor{\bsnm{Sumiya}, \binits{H.}},
\bauthor{\bsnm{Cappellaro}, \binits{P.}},
\bauthor{\bsnm{Knowles}, \binits{H.S.}},
\bauthor{\bsnm{Park}, \binits{H.}},
\bauthor{\bsnm{Lukin}, \binits{M.D.}}:
\batitle{Quantum metrology with strongly interacting spin systems}.
\bjtitle{Phys. Rev. X.}
\bvolume{10},
\bfpage{031003}
(\byear{2020})
\doiurl{10.1103/PhysRevX.10.031003}
\end{barticle}
\endbibitem

\bibitem[\protect\citeauthoryear{Wu}{}]{XY8ODMRTheory}
\begin{botherref}
\oauthor{\bsnm{Wu}, \binits{W.}}:
XY8-ODMR Theory.
in preparation.
\end{botherref}
\endbibitem

\bibitem[\protect\citeauthoryear{Dréau et~al.}{2011}]{Dreau2011}
\begin{barticle}
\bauthor{\bsnm{Dréau}, \binits{A.}},
\bauthor{\bsnm{Lesik}, \binits{M.}},
\bauthor{\bsnm{Rondin}, \binits{L.}},
\bauthor{\bsnm{Spinicelli}, \binits{P.}},
\bauthor{\bsnm{Arcizet}, \binits{O.}},
\bauthor{\bsnm{Roch}, \binits{J.-F.}},
\bauthor{\bsnm{Jacques}, \binits{V.}}:
\batitle{Avoiding power broadening in optically detected magnetic resonance of
  single nv defects for enhanced dc magnetic field sensitivity}.
\bjtitle{Phys. Rev. B}
\bvolume{84},
\bfpage{195204}
(\byear{2011})
\doiurl{10.1103/PhysRevB.84.195204}
\end{barticle}
\endbibitem

\end{thebibliography}

\end{document}


\title{Supplementary Information for ``A strongly interacting, two-dimensional, dipolar spin ensemble in (111)-oriented diamond"}
\author{L. B. Hughes, S. A. Meynell, W. Wu , S. Parthasarathy, L. Chen, Z. Zhang, Z. Wang, E. J. Davis, K. Mukherjee, N. Y. Yao, A. C. Bleszynski Jayich}
\date{\today}

\maketitle

\section{Dipolar interactions in 2D (111) versus (001) spin systems}\label{SIsec1}

\begin{figure}
\centering
\includegraphics[width = 120 mm]{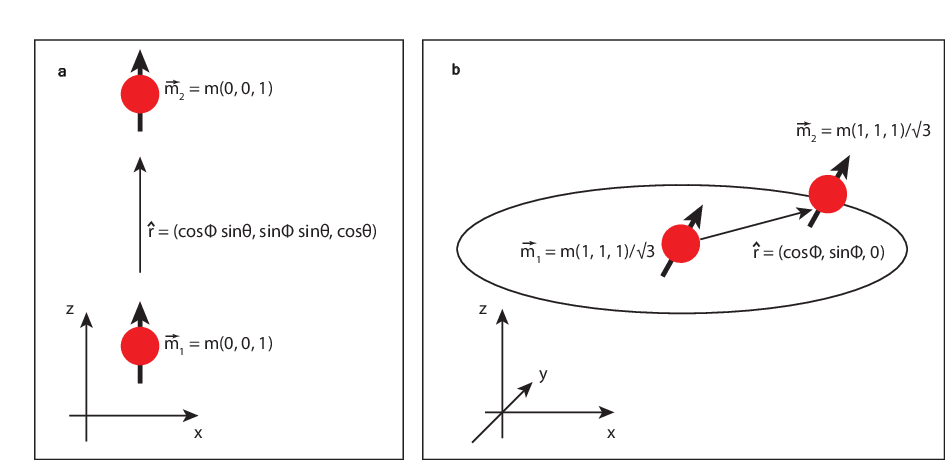}
\caption{\label{fig:SI_fig1} \textbf{Schematic of dipolar interactions.} a) The average value of $J$ in a 3D ensemble can be found by integrating Eqn.~\ref{eqn:3denergy} over $4\pi$ steradians. For a 3D ensemble, $\langle J\rangle = 0$. b) A two-dimensional ensemble may have non-zero or zero $\langle J\rangle$ depending on the orientation of the spins with respect to the plane. For 100 diamond, $\langle J\rangle = 0$, but for other orientations (such as 111), it is non-zero. Figure is reproduced from ref.~\cite{Meynell2023}.}
\end{figure}

Consider two classical magnetic dipoles in an isotropic 3D bath, $\vec{m}_1 = \vec{m}_2 = m(0,0,1)$ connected by an arbitrary vector, $\vec{r} = r\hat{r} = r(\cos\phi \sin\theta,\sin\phi \sin\theta, \cos\theta)$.
Figure~\ref{fig:SI_fig1}a shows a schematic illustration of this.
In the examples discussed here, every spin points along the same axis.
The energy of these two spins will be given by,
\begin{equation}\label{eqn:3denergy}
    U(r,\theta,\phi) = -\frac{\mu_0}{4\pi r^3}\left( 3(\vec{m}_1\cdot\hat{r})(\vec{m}_2\cdot\hat{r})-\vec{m}_1\cdot\Vec{m}_2\right) = f(r)(3\cos^2\theta-1),
\end{equation}
where $f(r) = -\mu_0m^2/4\pi r^3$ is introduced for notational simplicity. In addition, we define interaction strength \begin{equation}\label{eqn:2d111dependence}
    J(r, \theta) = f(r) (3 \cos^2 \theta - 1)
\end{equation}
Taking the integral over $4\pi$ steradians gives the average dipolar interaction strength
\begin{equation}
    \langle J \rangle = \int_0^{2\pi}\int_0^{\pi} f(r)(3\cos^2\theta - 1) \sin\theta~d\theta d\phi = 0.
\end{equation}
We next examine the case of a 2D ensemble of spins organized in the $xy$-plane.
For NVs in (001) diamond, all of these spins will point along the $[1,1,1]$ axis in the coordinate system shown in Figure~\ref{fig:SI_fig1} b.
Here, our dipolar energy will be given by
\begin{equation}\label{eqn:2d100dependence}
    J(r,\phi) = f(r)\left( (\cos\phi + \sin\phi)^2 - 1 \right),
\end{equation}
and integrating it over $4\pi$ steradians, we find
\begin{equation}\label{eqn:2d001Averaging}
\begin{split}
    \langle J \rangle & =\int_0^{2\pi}\int_0^{\pi} f(r) \left( (\cos\phi + \sin\phi)^2 - 1\right) \delta(\theta - \pi/2)\sin\theta~d\theta d\phi,\\
    & = \int_0^{2\pi} f(r)\left((\cos\phi + \sin\phi)^2 - 1\right)~d\phi \\ & =2f(r)\int_0^{2\pi} \cos\phi\sin\phi~d\phi= 0,\\
    & = 0.
    \end{split}
\end{equation}
We understand the averaging here to be a configurational averaging over an isotropic and uniform density confined to the plane where the Dirac delta function $\delta(\theta)$ is included to account for the spins being confined to the $xy$-plane.
In both the 3D isotropic case and the 2D (001) sample case, the dipolar interactions have a net zero value.
Interactions averaging away in 2D is only true for the special case of $001$ samples, where the NVs sit at a ``magic angle'' of $54.7356^{\circ}$ with respect to the plane.
For samples grown along the (111)-axis, the interactions \emph{do not} average away.

If we repeat the same calculation as Eqn.~\ref{eqn:2d001Averaging} but with the spins aligned along the plane normal (the $z$-axis), we get a dramatically different result.
In this case, $\vec{m}_1 = \vec{m}_2 - m(0, 0, 1)$, and
\begin{equation}
    J(r) = f(r)(3\cos^2\theta - 1) = -f(r).
\end{equation}
Integrating,
\begin{equation}\label{eqn:2d111Averaging}
\begin{split}
    \langle J \rangle & = \int_0^{2\pi}\int_0^{\pi} f(r) (3\cos^2 \theta - 1) \delta(\theta - \pi/2)\sin\theta~d\theta d\phi,\\
    & =  -f(r)\int_0^{2\pi}~d\phi \\
    & = -2\pi f(r) \\
    & \neq 0.
    \end{split}
\end{equation}
Thus, for a 2D layer of spins, all of which are pointed along the plane normal, the spin-spin interaction is not only on average non-zero but is also the same sign for every single pair of spins.
The dramatic differences between these cases suggest using dimensionality as a tool to access radically different regimes of dipolar coupling.

\section{Effects of doping time, temperature, and miscut on nitrogen incorporation}\label{SIsec2}

\begin{figure}
\centering
\includegraphics {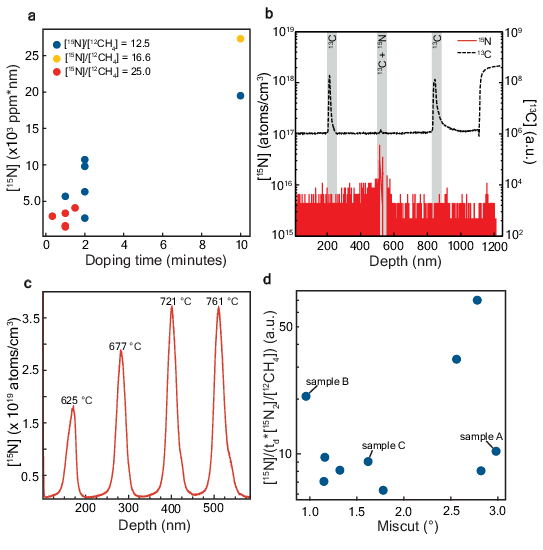}
\caption{\label{fig:SI_fig2} \textbf{Influence of doping time, temperature and miscut on nitrogen incorporation.} (a) Plot of nitrogen incorporation versus doping time, where samples are grouped by color according to the [$^{15}$N$_2$]/[$^{12}$CH$_4$] ratio.
(b) SIMS profile of an (001) sample co-doped with $^{13}$C before, during, and after introduction of $^{15}$N. The decreased $^{13}$C peak during nitrogen doping suggests a reduction in the diamond growth rate which may allow for increased nitrogen areal density.
(c) SIMS profile of a sample grown with four different doping temperatures, as indicated on the plot. In contrast to (001) growth, nitrogen shows a weakly increasing dependence with higher temperature.
(d) Scatterplot of samples showing SIMS-measured nitrogen density (normalized by t$_d$ and nitrogen gas content) versus polar miscut angle. No clear trend exists between miscut and nitrogen incorporation in (111) growth. }
\end{figure}

We find that increasing the nitrogen doping time retains two-dimensionality rather than creating a thicker layer, and as a result, the areal density of nitrogen increases with longer doping, which is shown in Fig.~\ref{fig:SI_fig2}a. 
This effect could be due to a non-constant diamond growth rate during doping; \textit{e.g.} if the growth rate slows significantly, then the dopant areal density can increase without the layer thickening.
This decelerated growth may be unexpected, given that small amounts of nitrogen typically result in an enhancement of the growth rate~\cite{Cao1996}.
However, given the extremely high ratio of [$^{15}$N$_2$]/[$^{12}$CH$_4$] used in our process, it is likely that the standard growth condition is disrupted in this way: the high amount of nitrogen changes the plasma and slows the diamond growth during doping.
We have seen evidence of this effect in a (001) sample grown under similar conditions (Fig.~\ref{fig:SI_fig2}b), where the introduction of a $^{13}$C co-dopant is used to independently track the diamond growth rate during nitrogen doping.
As shown in Fig.~\ref{fig:SI_fig2}b, when $^{15}$N is introduced, the $^{13}$C co-dopant peak is significantly reduced.
We hypothesize that a similar effect is occurring during (111) growth which allows for creation of high-density 2D nitrogen layers.

Temperature is known to affect the dopant incorporation efficiency in (001) diamond growth, where higher temperatures lead to greater desorption from the growing surface and thus lower density~\cite{Tallaire2015}.
In contrast, we find that higher doping temperature produces a weak increase in nitrogen incorporation, as can be seen in Fig.~\ref{fig:SI_fig2}c. 
This observation is similarly observed in ref.~\cite{Kato2007,Ozawa2017} and suggests that the thermal stability of dopants on the (111) surface is higher than on (001).
Fig.~\ref{fig:SI_fig2}d illustrates the nitrogen density measured with SIMS ([$^{15}$N]) versus miscut angle for ten (111) samples, where [$^{15}$N] has been normalized by the parameters that we found to affect incorporation: doping time and nitrogen:methane gas ratio.
For (001) growth, the substrate miscut sets the step-edge density which crucially impacts the nitrogen incorporation during step-flow growth, and therefore higher miscut is an important tuning knob for increasing nitrogen incorporation~\cite{Meynell2020}.
However, no clear trend is found amongst the (111) samples; thus, the nitrogen does not show a preference for sticking only at step edges sites, similar to the step-edge-independent growth mode we observed with carbon addition.
Altogether, our investigation of dopant incorporation in (111) growth has shown that the tuning knobs differ from (001), but we can still successfully create high-density, thin nitrogen layers via delta-doping.

\section{SIMS and DEER confirmation of two-dimensionality}\label{SIsec3}
\begin{figure}
\centering
\includegraphics {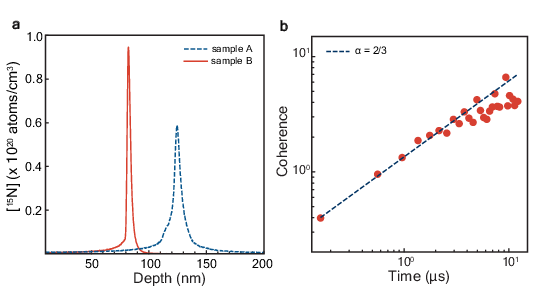}
\caption{\label{fig:SI_fig3} \textbf{Characterization of 2D nature via SIMS and DEER.} (a) SIMS depth profile of sample A (blue, dashed) and sample B (red, solid), showing thin confinement of the delta-doped layer. The FWHM of sample A = 5.8 nm and of sample B = 3.2 nm and the areal densities are A = 1500 ppm$\cdot$nm and B = 1700 ppm$\cdot$nm, respectively. 
(b) Log-log plot of the DEER decay in sample A showing good agreement with an $\alpha$ = 2/3 stretch exponent, indicating that the NVs probe a 2D P1 center bath. }
\end{figure}

SIMS depth profiles of sample A and B show thin delta-doped layers and similar nitrogen content in the two samples, which were grown simultaneously (Fig.~\ref{fig:SI_fig3}a). 
The FWHM of sample A = 5.8 nm and of sample B = 3.2 nm. 
The broadening of sample A data may be due to increased surface roughness (see Fig. 2b in the main text) or charging-induced drifts that were observed during data collection.
The SIMS data is integrated over a range of 3*$\sigma$ from the peak center to estimate the nitrogen areal density: sample A = 1500 ppm$\cdot$nm and B = 1700 ppm$\cdot$nm, and dividing by the layer thickness yields the volume density of sample A = 260 ppm and B = 520 ppm.
These densities are notably smaller than the sample shown in Fig. 1b, because they were grown with 1/10th of the doping time.
Fig.~\ref{fig:SI_fig3}b shows a log-log plot of the P1 DEER decay in sample A. 
The dashed line indicates a stretch exponent of $\alpha$ = 2/3, indicative of a 2D nitrogen spin bath~\cite{Davis2023}.
This data supports our XY-8 characterization of a 2D NV spin bath in sample A.

\section{Influence of miscut on grown-in NV density}\label{SIsec4}
\begin{figure}
\centering
\includegraphics {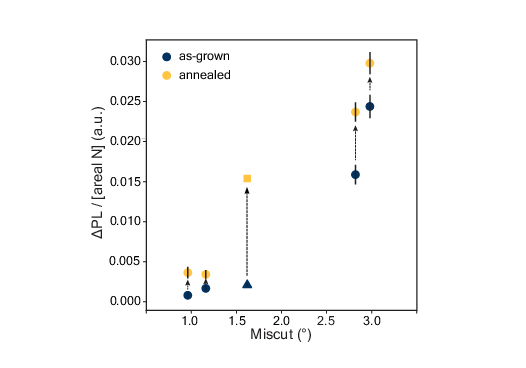}
\caption{\label{fig:SI_fig4} \textbf{Influence of miscut on NV density.} Plot of $\Delta$PL, a proxy for NV density, divided by the areal density of nitrogen extracted from SIMS versus substrate miscut.
Blue data represents as-grown NV concentrations, and yellow shows enhancement after annealing.
$\Delta$PL values are measured under 100 $\micro$W excitation, and all circular points are measured on the same confocal setup; square and triangle points are measured on a different setup and may be prone to increased error due to differences in the collection optics.
Error bars represent the standard error of the mean of $\Delta$PL measurements taken on different locations within the diamond sample.
}
\end{figure}

Here we present NV incorporation during growth versus a wide range of substrate miscuts, finding that the trend of increasing NV density with higher miscut holds across many samples.
Fig.~\ref{fig:SI_fig4} shows $\Delta$PL, a proxy for NV density, normalized by the areal density of nitrogen extracted from SIMS and plotted against miscut angle for samples as-grown (blue) and directly after annealing (yellow). 
The highest and lowest miscut samples plotted here are sample A and B from the main text, respectively.
The $\Delta$PL estimation (PL difference between the $m_s=0$ and $\pm1$ states during a Rabi measurement) is critically sensitive to the laser power used~\cite{Hughes2023}, and therefore all of the circular data points are taken on the same confocal setup under 100 $\micro$W excitation.
The triangle and square data represent measurements performed on a different confocal setup with different collection efficiencies which could affect the estimation of $\Delta$PL.
In ref.~\cite{Hughes2023}, $\Delta$PL is converted to a quantitative NV density via calibration with a single NV; however, even the lowest miscut (111) sample contains greater than single NV-level densities and hence we cannot convert $\Delta$PL directly to a quantitative density.
Nevertheless, Fig.~\ref{fig:SI_fig4} highlights how the substrate miscut can be used as a tuning parameter to control the amount of NV centers incorporated during (111) growth.

The physical mechanism driving the dependence of NV incorporation during growth on substrate miscut could have a few origins. 
Prior works have observed increased NV density with lower growth temperature, attributed to greater vacancy incorporation resulting from decreased crystalline quality~\cite{Ozawa2017,Tallaire2015}. A similar hypothesis may support our findings: higher miscut substrates have a more disordered surface that could be amenable to greater vacancy incorporation.
The faster growth rates associated with higher miscut may also result in greater vacancy incorporation due to decreased crystalline quality.
Further investigation is needed to elucidate the precise mechanism of NV formation during growth with varying miscut.

\section{Sensing Properties of (111) delta-doped ensembles}\label{SIsec5}
\subsection{Sensitivity Estimation}
\begin{table}[]
    \centering
    \begin{tabular}{|m{2cm}|m{1.5cm}|m{1.5cm}|m{1.5cm}|m{1.5cm}|m{1.5cm}|}
    \hline
         & sample A, as-grown & sample A, TEM (4e19 e$^{\text{–}}$/cm$^2$) & sample B, as-grown & sample B, TEM (7e18 e$^{\text{–}}$/cm$^2$) & sample B, projected\\ \hline\hline
        Sequence type & Hahn Echo & XY-8 & Hahn Echo & Hahn Echo & Hahn Echo \\ \hline
        T$_2$ ($\mu$s) & 2.7 & 48 & 4.9 & 65 & 65 \\ \hline
        Sensing time, $\tau$ ($\mu$s) & 1.35 & 24 & 2.5 & 32.5 & 32.5 \\ \hline
        Overhead time, t$_m$ ($\mu$s) & 20.4 & 20.4 & 20.4 & 20.4 & 4.4 \\ \hline
        Readout efficiency, $C_{eff}$ & 0.017 & 0.0062 & 0.0079 & 0.00098 & 0.043 \\ \hline
        Number of NVs & 20 & 135 & 2 & 10 & 1 \\ \hline
        Volume, V ($\mu$m$^3$) & 0.0015 & 0.0015 & 0.0015 & 0.0015 & 0.0015 \\ \hline
        NV density, $\rho$ (ppm) & 0.75 & 0.83 & 0.1 & 0.5 & 0.5 \\ \hline
        $\rho$*T$_2$ & 0.36 & 0.88 & 0.063 & 4.5 & 4.5 \\ \hline
        Volume-normalized sensitivity, $\eta_{AC}$ (nT µm$^{3/2}$ Hz$^{-1/2}$) & 5.85 & 9.83 & 19.5 & 0.81 & 0.153 \\ 
    \hline
    \end{tabular}
    \caption{\textbf{Measurement parameters used to calculate sensitivity for samples A and B.} We note that the T$_2$ improves in the TEM spots (and quite significantly for sample B), which is likely related to P1-NV conversion or charge destabilization of other noise-producing defects, and hence the best sensitivity is found in the low-dose (7$\times$10$^{18}$ e$^{\text{–}}$/cm$^2$) TEM spot of sample B.}
    \label{tab:tab1}
\end{table}

The AC sensitivity of our (111) delta-doped samples is calculated according to Eq.~\ref{eq:eq2}~\cite{Barry2020} and the values given in table~\ref{tab:tab1}.  
\begin{equation}
\label{eq:eq2}
    \eta_{echo}\approx\frac{\pi}{2}\frac{\hbar}{g_{e}\mu_{B}}\frac{{e^{(\tau/T_2)^p}}}{C_{eff}\sqrt{N}}\frac{\sqrt{t_m + \tau}}{\tau}
\end{equation}
Where $\tau$ is the sensing time, $p=2/3$ for a 2D system, $t_{m}$ is the measurement overhead time, $C_{eff} = C \sqrt{\frac{PL \times t_R}{N}}$ where $C$ is the contrast as defined in ref.~\cite{Barry2020}, $PL$ is the photon count rate, $t_R$ is the readout time, and $N$ is the number of NVs. 
For a volume-normalized sensitivity, the scaling of sensitivity with $\frac{1}{\sqrt{V}}$ is removed by replacing $N$ with the 3D NV density $\rho_{3D}$.

In the main text, we use empirically determined coherence to calculate sensitivity, but some care is required to interpret the outlook for improvements. 
For large densities, $T_2$ is limited by dipolar interactions and sensitivity ceases to scale favorably with density in the naive Hahn Echo protocol. 
To see this, we further approximate the sensitivity in the limit of zero overhead time $t_m = 0$ and optimally chosen sensing time $\tau = \alpha(p) T_2$, where $\alpha$ is an order 1 factor that depends on $p$:
\begin{equation}
    \eta_{echo}\approx\frac{\pi}{2}\frac{\hbar}{g_{e}\mu_{B}}\frac{{e^{\alpha^p}}}{C_{eff}\sqrt{\alpha}\sqrt{N T_2}} \sim \frac{1}{\sqrt{N T_2}}
\end{equation}
As derived in the supplemental material for ref.~\cite{Hughes2023}, for a dipolar-limited ensemble of dimension $D$, $\frac{1}{T_2^{D/3}} \sim \rho_D J^{D/3}$, where $\rho_D$ is the density in the appropriate units for dimension $D$. 
Thus, for a 3D ensemble, $\rho_{3D} T_2$ is constant, whereas for a 2D ensemble, $\rho_{2D} T_2^{2/3} = w \ \rho_{3D} T_2^{2/3}$ is constant, where $w$ is the layer thickness. 
Using $N = \rho_{3D} V = \rho_{2D} A$ (where $A$ is the area for a 2D ensemble), we find that volume-normalized Hahn Echo sensitivity $\eta_{echo}^{3D} \sqrt{V}$ is independent of density for a 3D dipolar-limited ensemble, but is even less favorable for a 2D dipolar-limited ensemble: $\eta_{echo}^{2D} \sqrt{V} \sim \rho_{3D}^{3/4} w^{1/4}$ or in terms of the more natural ``area-normalized" sensitivity, $\eta_{echo}^{2D} \sqrt{A} \sim \rho_{2D}^{1/4}$. 
In order to reap the gains of increased sensor density in 2D, one has to decouple dipolar interactions, which cannot be done effectively via conventional XY8-based sensing protocols. 
However, as demonstrated in ref.~\cite{Zhou2020}, this barrier is relatively straightforward to overcome by switching to DROID based sensing. 
The condition for when sensitivity based on the DROID protocol is lower than sensitivity based on the XY8 protocol is $T_{2,DROID} > 3 T_{2,XY8}$. 
For most clean dipolar-limited ensembles we expect this to be achievable and for DROID-based sensing to be the optimal method to exploit denser 2D samples for sensing, though the practical limits of this sensing protocol (aside from $T_1$) are not yet clear. 

\subsection{Magnetic imaging demonstration}
\begin{figure}
\centering
\includegraphics {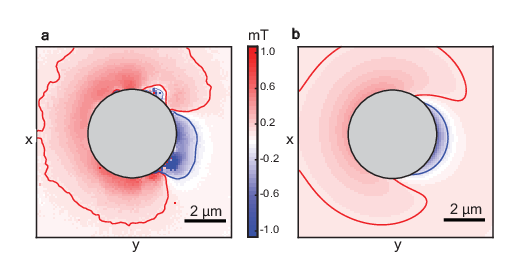}
\caption{\label{fig:SI_fig5} \textbf{Magnetic imaging with (111) NV ensembles.} (a) Plot of the magnetic field produced by an Fe nanoparticle deposited on the diamond surface and encapsulated in type F objective oil. (b) Dipole simulation for the field produced by a 2 µm-diameter iron particle with a magnetic moment of 2.216 µB/atom. Field contours at $\pm$ 0.1 mT are indicated by red and blue lines in both plots. A gray circle is drawn over the center of the particle where NV photoluminescence is quenched and ODMR contrast is depleted.}
\end{figure}

We next demonstrate magnetic field imaging of a micrometer-scale Fe nanoparticle deposited atop diamond sample A with $\sim$ 120 nm-deep, grown-in (unirradiated) NV centers.
ODMR is performed over a 10x10 µm$^{2}$ region of NV centers, taking 100x100 pixels and sweeping the frequency from 2750-2850 MHz.
The resulting spatial magnetic field plot is shown in Fig.~\ref{fig:SI_fig5}a, with contour lines indicated at $\pm$0.1 mT.
A gray circle is drawn over the center of the particle because in this region the ODMR contrast is depleted and the NV photoluminescence is quenched.
Fig.~\ref{fig:SI_fig5}b shows a simulated dipole field due to a 2 µm-diameter Fe particle, assuming a magnetic moment of 2.216 µB/atom, and a 2D ensemble of (111) NV sensors 120 nm away.
The dipole orientation is simulated with a polar angle $\theta$ = 162$\degree$ (into the page) and azimuthal angle $\phi$ = 164$\degree$ (in the x-y plane).
Similar dipole contours are drawn at $\pm$ 0.1 mT. 
The simulated field is $~$100x smaller than expected, which could be due to increased standoff from the oil or smaller than expected particle radius.
This demonstration shows that our high-density (111)-aligned NVs provide a platform for applications in ensemble magnetometry.

\section{Summary of samples}\label{SIsec6}
Table~\ref{tab:tab2} shows each sample studied in this work and the relevant characteristics. 
Sample C is a delta-doped (111) sample with SIMS data shown in Fig.~1b.
Sample D is a delta-doped (001) sample, also referred to as ``sample S1" in ref.~\cite{Davis2023}.
We note that sample D represents the highest density (001) sample that we have produced to date and its creation has not yet been reproducible, unlike with (111) samples.
We report NV/P1 ratios as a metric for NV$^-$ creation efficiency, since SIMS measures all nitrogen-related defects including those which cannot be converted to NV$^-$\cite{Hughes2023}.

\begin{table}[]
    \centering
    \begin{tabular}{|m{3cm}|m{1.5cm}|m{1.5cm}|m{1.5cm}|m{1.5cm}|}
    \hline
         & sample A & sample B & sample C & sample D \\ \hline\hline
        crystal orientation & (111) & (111) & (111) & (001) \\ \hline
        miscut, $\theta$ (degrees) & 3.0 & 0.96 & 1.6 & 1.1 \\ \hline
    [$^{15}$N$_2$]/[$^{12}$CH$_4$] ratio & 25 & 25 & 12.5 & 25 \\ \hline
        doping time (minutes) & 1 & 1 & 10 & 10 \\ \hline
        SIMS nitrogen layer thickness (nm) & 5.8 & 3.2 & 17 & 8.8 \\ \hline
        SIMS nitrogen areal density (ppm$\cdot$nm) & 1500 & 1700 & 19000 & 200 \\ \hline
        SIMS nitrogen volume density (ppm) & 260 & 520 & 1100 & 23 \\ \hline
        As-grown NV density (ppm$\cdot$nm) & 4.5 & 0.65 & -- & -- \\ \hline
        As-grown DEER P1 density (ppm$\cdot$nm) & 126 & 6 & -- & -- \\ \hline
        As-grown NV/P1 ratio & 0.04 & 0.11 & -- & -- \\ \hline
        TEM irradiation dose (e$^{\text{-}}$/cm$^2$) & 4e19 & -- & 6e19 & 7e20 \\ \hline
        TEM DEER P1 density (ppm$\cdot$nm) & 111 (2e19 dose) & -- & 64 & 85 \\ \hline
        TEM NV density (ppm$\cdot$nm) & 10 & -- & 8 & 19 \\ \hline
        TEM NV/P1 ratio & 0.09 & -- & 0.13 & 0.22 \\ \hline
    \hline
    \end{tabular}
    \caption{\textbf{Details of samples studied in this work.} Dashed lines indicate unmeasured quantities. ``As-grown" measurements are taken on the unirradiated regions of the diamond, but the samples have been through an annealing process before characterization.}
    \label{tab:tab2}
\end{table}

\section{XY8-ODMR details}\label{SIsec7}
\subsection{Theory of measurement}
The type of dynamical decoupling determines how robust this sequence is to pulse error, where the ODMR peaks appear in frequency space, and in non-ideal cases how much disorder remains.
The XY8-ODMR sequence applies a weak probe $\pi$ pulse during an XY8 $\pi$ pulse train where we scan the microwave frequency of the weak probe pulse to detect the resonance.
Note that the initial state is prepared at either $0$ or $-1$ (or $+1$), which is different from the conventional XY8 sequence for dynamical decoupling, where the initial state is a coherent superposition of $0$ and $-1$ (or $+1$).
The Hamiltonian in the lab frame can be written as
\begin{equation}
\begin{aligned}
    H =& (\omega_0 + \omega_{\mathrm{disorder}} + \omega_{NV}(t) )S_z \\ 
    &  + 2[\Omega_1'(t)\sin(\omega_0 t) + \Omega_2'(t)\cos(\omega_0 t) + \Omega \sin((\omega_0+\omega_d) t + \phi)] S_y,
\end{aligned}
\end{equation}
where $\omega_0$ is the bare splitting of the NV, $\omega_{\mathrm{disorder}}$ is the static disorder, $\omega_{NV}(t)$ is the NV-NV interaction, which is time-dependent because the $\pi$ pulse train keeps alternating NV's quantum state between $0$ and $-1$, \textit{i.e.},
\begin{align}
    \omega_{NV}(t) &=
    \begin{cases}
        \omega_{NV} & \text{if } n \text{\ is even (NV is in the $-1$ state)},\\
        0 & \text{otherwise (NV is in the $0$ state)},
    \end{cases}
\end{align}
$\phi$ is the relative phase between the fast and slow drive.
The slow ODMR  driving is of a microwave frequency $\omega_0 + \omega_d$ and a Rabi frequency $\Omega$.
The fast XY8 driving is of a microwave frequency $\omega_0$ (on resonance) and a Rabi frequency $\Omega_1'(t)$ and $\Omega_2'(t)$, which corresponds to the amplitude of the X and Y pulses.
Here, we only keep the Ising interaction between NVs which contributes to the spectrum and ignore the flip-flop interaction. 

In order to analyze the behavior of such pulse sequence, we can transform the Hamiltonian into the toggling frame, the new spin operators $\tilde{S}_i$ are related to the ones $S_i$ in the original frame by $S_i = c_i(t)\tilde{S}_i$, where $i = x, y, z$.
The Hamiltonian in the toggling frame is 
\begin{equation}
    H = [\omega_{\mathrm{disorder}} + \omega_{NV}(t) ]c_z(t)\tilde{S}_z + \Omega [\cos(\omega_d t + \phi) c_x(t) \tilde{S}_x + \sin(\omega_d t + \phi) c_y(t) \tilde{S}_y],
\end{equation}
where the coefficients are  
\begin{align}
    c_x(t) &=
    \begin{cases}
        1 & \text{if } t < \tau_p, 3\tau_p<t<4\tau_p, 6\tau_p < t < 8\tau_p\\
        -1 & \text{otherwise},
    \end{cases}\\
    c_y(t) &=
    \begin{cases}
        1 & \text{if } 2\tau_p < t < 5\tau_p, 7\tau_p<t<8\tau_p\\
        -1 & \text{otherwise},
    \end{cases}\\
    c_z(t) &=
    \begin{cases}
        1 & \text{if } n \text{\ is even},\\
        -1 & \text{otherwise}.
    \end{cases}
\end{align}
The first term in the Hamiltonian gives the disorder-free spectrum which is only limited by NV-NV interaction.
Because $c_z(t)$ alternates its sign during the pulse sequence, it echoes out the static disorder $\overline{\omega_{\mathrm{disorder}}c_z(t)} = 0$ on average while the NV-NV interaction survives $\overline{\omega_{NV}(t)c_z(t)} = \omega_{NV}/2$.
The second term in the Hamiltonian is the weak probe pulse. Due to the time dependence of the coefficient $c_x(t)$ and $c_y(t)$, we perform the Fourier transform for the weak probe pulse, and get different frequency components, resembling a frequency comb with a comb tooth spacing $1/8\tau_p$.
In this experiment, we use the comb tooth $n=-2$, i.e., the driving frequency is $\omega_{\mathrm{probe}} = \omega_d - 2/(8\tau_p)$.
As a result, the resonance is detected when $\omega_{\mathrm{probe}} = \omega_{NV}/2$, i.e., $\omega_d= \omega_{NV}/2 + 2/(8\tau_p)$.
In the experiment, the spectrum is observed at the microwave frequency $\omega_0+\omega_d$ = $\omega_0 + \omega_{NV}/2 + 2/(8\tau_p)$, and the shape of the spectrum directly maps out the random distribution of $\omega_{NV}$. A theoretical framework and more details of XY8-ODMR protocols can be found in a separate work in preparation \cite{XY8ODMRTheory}.  

\subsection{Simulation of spectrum}
Here we summarize a model for the dipolar coupling distribution and how the theoretical curves in Fig.~4 were constructed.
Each NV in our ensemble is subject to a mean magnetic field produced by other NVs nearby.
The strength and direction of this mean field is determined by the distances to other NVs (set by NV density), relative angle between NVs (0$\degree$ for (111) or 54.7$\degree$ for (001) orientation), and spin orientation (typically optically pumped into $m_s$ = 0).
One effect captured by this model, visualized in the data for (111) Sample C in Fig.~1 is that the mean field produced by the surrounding NV bath is positive or negative relative to the external B field depending on $m_s$, causing a change in the asymmetry depending on spin initialization. 

Importantly, our ensemble measurements include many positional configurations due to imperfect spin initialization fidelity and because the confocal spot is much larger than typical length scales over which NV-NV interactions are significant (\textit{i.e.}, different parts of the confocal spot can be treated as independently sampled positional configurations).
To simulate a mean field for an NV, $K$ spins were randomly positioned in a 400 nm $\times$ 400 nm $\times$ $t$ nm prism centered at the NV's location (defined to be the origin), with the (111) direction defined to point along the $t$ nm edge. $t$ was fixed in accordance with the parameters in Table~\ref{tab:tab2}.
For the curves shown in Fig.~4 of the main text, we allowed $K$ to vary and found better alignment with the data if the NV density used was $70\%$ of the density listed in line 12 of Table~\ref{tab:tab2}, which is consistent with XY8-based density estimation likely overestimating the density by a small factor due to the presence of dynamical disorder.
The total field felt by the NV was calculated as $B = \sum_i \frac{J}{r_i^3} s_i f(r_i, \theta_i, \phi_i)$, where $J = 2 \pi \times 52$ MHz $\mathrm{nm^3}$, $r_i, \theta_i, \phi_i$ are the positions of spin $i$ in cylindrical coordinates, $f(r_i, \theta_i, \phi_i) = 3 \cos^2(\theta_i) - 1$ for a (111) ensemble or $(\cos(\phi_i)\sin(\theta_i) + \cos(\phi_i)\sin(\theta_i) + \cos(\theta_i))^2 - 1$ for an (001) ensemble, and $s_i = \pm \frac{1}{2}$ is the value of the spin in the reduced spin-1/2 qubit basis (e.g., for 100\% spin initialization fidelity, $s_i = \frac{1}{2}$ for all spins).
We note that this more complicated angular dependence for (001) ensembles reduces to the formulas in Eqs.~\ref{eqn:2d111dependence} and \ref{eqn:2d100dependence} for an infinitely thin layer.
The full histogram of mean fields presented in Fig.~1 was constructed by calculating the B-field as above 100,000 times using the experimentally measured density and thickness from Table~\ref{tab:tab2}, Sample C.

In order to construct a more realistic theoretical spectrum for the XY8-ODMR measurement presented in the main text, we introduced two broadening mechanisms.
First, in all cases we convolve the histogram with a lorentzian of linewidth $2 \alpha \Omega$ (where $\Omega$ is the probe ODMR rabi frequency, and $\alpha = 1 - 1.2$ is a small order 1 factor) to approximate the effects of both power broadening (2 $\Omega$) and any disorder ($\alpha$) that remains after dynamical decoupling.
Second, we simulate the effect of imperfect initialization fidelity by assuming 33\% of the spins in the simulation are pointing in the opposite direction with the same magnitude field. 
This ``initialization fidelity" parameter likely accounts for imperfect spin initialization at the beginning of the sequence as well as imperfect pi pulse fidelity during the sequence.

\subsection{Effect of power broadening}
\begin{figure}
\centering
\includegraphics{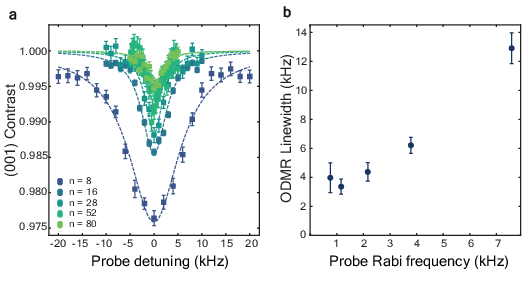}
\caption{\label{fig:SI_fig6} \textbf{Linewidth saturation of XY8-ODMR on (001) ensembles.} (a) Various XY8-ODMR spectra for different XY8 repetitions. Since the probe ODMR pulse duration is matched to that of total XY8 pulse duration, we effectively reduce the probe ODMR Rabi frequency as we increase XY8 repetitions. The spectra are fitted with Lorenztian distributions and normalized to the background of the fits. (b) Extracted linewidth as a function of Rabi frequency. We notice the ODMR linewidths become roughly unchanged when the probe Rabi frequency is reduced below 1 kHz.}
\end{figure}

The dipolar-interaction limit is clearly observed for (111) ensembles as evidenced by the asymmetry in the XY8-ODMR spectra.
However, characterizing this limit for (001) ensembles is more challenging since the true lineshape could be masked by broadening mechanisms.
One of the main sources of broadening is due to the laser power of the probe ODMR pulse as well as imperfect initialization.
In order to remove this source, we increase the probe ODMR pulse duration (\textit{i.e.}, by reducing probe ODMR Rabi frequency) \cite{Dreau2011}.
At small enough Rabi frequencies $\leq$ 1 kHz, we observe saturation of the spectral linewidth, as is shown in Fig.~\ref{fig:SI_fig6}. 
This saturation confirms that we have reached the dipolar-limited regime for the (001) sample and can confidently quantify the symmetry of the lineshape. 


\bibliography{111_SI}